\documentclass[aps,prl,twocolumn,superscriptaddress]{revtex4-2}

\usepackage[T1]{fontenc}
\usepackage[utf8]{inputenc} 

\usepackage{amsmath}
\usepackage{amsfonts}
\usepackage{amssymb}
\usepackage{graphicx}
\usepackage{bm}
\usepackage{color}
\usepackage{multirow}
\usepackage{lineno}
\usepackage{dsfont}
\usepackage{siunitx} 
\usepackage{lipsum}
\usepackage{array}
\usepackage{newtxtext,newtxmath}

\usepackage{verbatim}
\usepackage{mathtools}
\usepackage{empheq}

\usepackage[normalem]{ulem}  
\usepackage{cancel}

\definecolor{darkgreen}{rgb}{0.09, 0.55, 0.3}

\definecolor{darkred}{rgb}{0.8, 0.10, 0.1}

\newcommand{\bea}{\begin{eqnarray}}
\newcommand{\eea}{\end{eqnarray}}
\newcommand{\be}{\begin{equation}}
\newcommand{\ee}{\end{equation}}
\newcommand{\bes}{\begin{equation*}}
\newcommand{\ees}{\end{equation*}}
\newcommand{\bi}{\begin{itemize}}
\newcommand{\ei}{\end{itemize}}

\DeclareSymbolFont{usualmathcal}{OMS}{cmsy}{m}{n}
\DeclareSymbolFontAlphabet{\mathcal}{usualmathcal}

\newcolumntype{P}[1]{>{\centering\arraybackslash}p{#1}}
\newcommand{\pr}[1]{\ensuremath{\left[#1\right]}} 
\newcommand{\pc}[1]{\ensuremath{\left(#1\right)}}

\newcommand{\ket}[1]{\ensuremath{\left\vert#1\right\rangle}}

\newcommand{\av}[1]{\ensuremath{\left\langle#1\right\rangle}} 
\definecolor{bluedarkRL}{rgb}{0,0.3,0.79}

\definecolor{burgundy}{rgb}{0.5, 0.0, 0.13}
\definecolor{denim}{rgb}{0.08, 0.38, 0.74}
\definecolor{midnightgreen}{rgb}{0.0, 0.29, 0.33}
\definecolor{sienna}{rgb}{0.53, 0.18, 0.09}
\definecolor{sacramentostategreen}{rgb}{0.0, 0.34, 0.25}

\DeclareSIUnit\gauss{G}
\definecolor{Green}{rgb}{0,0.6,0.4}

\usepackage{xcolor}
\usepackage[%
colorlinks=true,
urlcolor=bluedarkRL,
linkcolor=bluedarkRL,
citecolor=bluedarkRL
]{hyperref}
\hypersetup{
colorlinks=true,
linkcolor=bluedarkRL,
citecolor=bluedarkRL,
urlcolor=black}

\begin{document}

\title{
Coexisting Regular and Chaotic Dynamics in the Dysprosium Feshbach Spectrum}
\author{Julie Veschambre}
\affiliation{Laboratoire Kastler Brossel, Coll\`ege de France, CNRS, ENS-Universit\'e PSL, Sorbonne Universit\'e, 11 Place Marcelin Berthelot, 75005 Paris, France}
\author{Alexandre Journeaux}
\affiliation{Laboratoire Kastler Brossel, Coll\`ege de France, CNRS, ENS-Universit\'e PSL, Sorbonne Universit\'e, 11 Place Marcelin Berthelot, 75005 Paris, France}
\author{Maxime Lecomte}
\affiliation{Laboratoire Kastler Brossel, Coll\`ege de France, CNRS, ENS-Universit\'e PSL, Sorbonne Universit\'e, 11 Place Marcelin Berthelot, 75005 Paris, France}
\author{Alice Belmon}
\affiliation{Laboratoire Kastler Brossel, Coll\`ege de France, CNRS, ENS-Universit\'e PSL, Sorbonne Universit\'e, 11 Place Marcelin Berthelot, 75005 Paris, France}
\author{Ethan Uzan}
\affiliation{Laboratoire Kastler Brossel, Coll\`ege de France, CNRS, ENS-Universit\'e PSL, Sorbonne Universit\'e, 11 Place Marcelin Berthelot, 75005 Paris, France}
\author{In\`es de Verdelhan}
\affiliation{Laboratoire Kastler Brossel, Coll\`ege de France, CNRS, ENS-Universit\'e PSL, Sorbonne Universit\'e, 11 Place Marcelin Berthelot, 75005 Paris, France}
\author{Patricia Christina Marques Castilho}
\affiliation{Instituto de Física de São Carlos, Universidade de São Paulo, CP 369, 13560-970 São Carlos, Brazil} 
\author{Jakub Zakrzewski}
\affiliation{Instytut Fizyki Teoretycznej, Wydzia\l{} Fizyki, Astronomii i Informatyki Stosowanej, Uniwersytet Jagiello\'nski, {\L}ojasiewicza 11, PL-30-348 Krak\'ow, Poland}
\affiliation{Mark Kac Complex Systems Research Center, Jagiellonian University in Krak\'ow, PL-30-348 Krak\'ow, Poland}\author{Jean Dalibard}
\affiliation{Laboratoire Kastler Brossel, Coll\`ege de France, CNRS, ENS-Universit\'e PSL, Sorbonne Universit\'e, 11 Place Marcelin Berthelot, 75005 Paris, France}
\author{Raphael Lopes}
\email{raphael.lopes@lkb.ens.fr}
\affiliation{Laboratoire Kastler Brossel, Coll\`ege de France, CNRS, ENS-Universit\'e PSL, Sorbonne Universit\'e, 11 Place Marcelin Berthelot, 75005 Paris, France}

\begin{abstract}
Strongly dipolar gases, such as dysprosium, erbium and thulium, exhibit dense Feshbach spectra whose level statistics have been associated with quantum chaos arising from couplings among many molecular channels. 
Here, we combine a precise calibration of the Feshbach spectrum of $^{162}$Dy with spectroscopic measurements of the differential magnetic moments of bound states associated with more than 80 resonances between 0 and 30~G. 
These magnetic moments provide an eigenstate-sensitive probe of the molecular states underlying the resonance spectrum. 
We find that the level statistics are not uniform: resonances associated with states near the center of the magnetic-moment distribution display enhanced level repulsion, whereas those near the lower edge remain close to Poisson statistics. 
Our results reveal hidden structure within the chaotic dysprosium Feshbach spectrum and identify molecular-state composition as a key ingredient in the emergence of quantum chaos in strongly dipolar scattering.
\end{abstract}

\maketitle

Strongly dipolar gases, such as dysprosium and erbium, provide a platform for probing many-body phases and lattice models, including droplet clusters, supersolidity, and extended Bose- and Fermi-Hubbard models with long-range interactions~\cite{Baier2016,Su2023,Chomaz2023,Biagioni2024}.
At the same time, the long-range and anisotropic dipole-dipole interaction considerably complicates the description of few-body collisions by coupling spin and orbital degrees of freedom, as illustrated by dipolar relaxation~\cite{Lecomte2025,Lafforgue2025} and the Einstein--de Haas effect~\cite{Gawryluk07, Swi2011, Matsui2026}.
This complexity is also reflected in the dense Feshbach spectra observed in ultracold samples of dysprosium, erbium, and thulium, for which a complete microscopic assignment, or even a statistical prediction of resonance positions, remains challenging~\cite{Baumann2014,Frisch2015,Maier2015,Khlebnikov2019}.

The first observations of dense~\cite{Baumann2014} and chaotic level statistics in these resonance spectra~\cite{Frisch2014,Maier2015,Maier2015b} stimulated substantial theoretical efforts to identify the microscopic mechanisms at their origin~\cite{Kotochigova2011,Petrov2012,Kotochigova2014,Makrides2018,Augustovivcova2018,Mccann2021}.
They also raised broader questions about the meaning of quantum chaos in strongly anisotropic molecular systems, and about which experimental observables are sufficient to characterize it~\cite{Mur2015,Roy2017,Casal2021}. Those recent studies supplement earlier treatments of quantum chaos in atomic and molecular systems, mainly by semiclassical methods (see, e.g. \cite{Heller75, Delande, Bluemel} and references therein).

Theoretical work has emphasized that Feshbach-resonance statistics reflect the near-threshold molecular spectrum, but that additional eigenstate-sensitive observables are needed to characterize the mixing and composition of the underlying molecular states~\cite{Augustovivcova2018, Mccann2021}.
Complementary diagnostics include the parametric evolution of energy levels, avoided crossings, and level curvatures~\cite{Zakrzewski1993, Zakrzewski1993b, Weidenmuller2009}, for reviews see \cite{Haakebook, Stoeckmann}.
This motivates the search for additional observables capable of resolving more detailed structure in strongly dipolar Feshbach spectra.

More generally, chaotic spectra need not be statistically homogeneous: regular and chaotic components can coexist, leading to mixed statistics~\cite{Rosenzweig1960,Berry1984,Delande, Prosen1999,Batistic2013,Batistic2019, Kosicki2020}.
A global level-spacing analysis can then obscure hidden structure, since only part of the spectrum may show strong level repulsion and random-matrix-like statistics~\cite{Gubin2012,Giraud2022,Zakrzewski2023,Yan2025}.
The challenge is to identify an experimentally accessible observable that can partially resolve these components.

\begin{figure*}[t!]
    \centering
    \includegraphics[width=17cm]{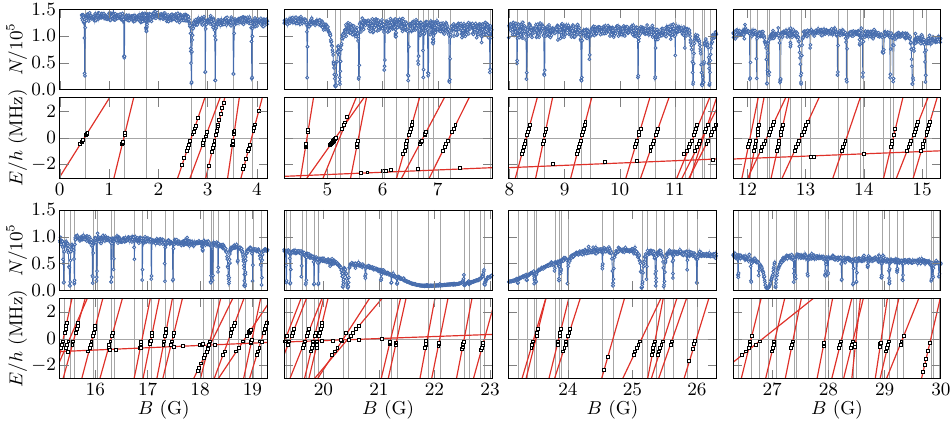}
\caption{Overview of the $^{162}$Dy Feshbach spectrum and differential-magnetic-moment spectroscopy. 
The figure is divided into eight magnetic-field windows covering 0 to 30~G.
For each window, the upper panel shows the atom-loss spectrum used to locate Feshbach resonances, measured with a magnetic-field resolution of 6~mG; the vertical gray lines mark the extracted resonance positions.
The lower panel shows the measured molecular binding energies $E/h$ as a function of magnetic field. 
Red lines are local linear fits used to extract the differential magnetic moments $\delta\mu$ of the associated molecular states.
We identify 128 resonances and determine $\delta \mu$ for 83 of them.
}
    \label{figFFRscan}
\end{figure*}

In this Letter, we probe not only the positions of Feshbach resonances, but also the differential magnetic moments of the associated molecular states. These moments, whose statistical properties had not previously been investigated experimentally, provide important information on the magnetic character of the bound states and on their composition in the relevant molecular basis, including spin, rotational, and vibrational degrees of freedom.
More importantly, they reveal statistically distinct components of the Feshbach spectrum with different degrees of level repulsion.
We show that the differential magnetic moment acts as an experimentally accessible classifier: resonances associated with states near the center of the measured distribution display enhanced level repulsion, whereas those near its lower edge remain close to Poisson statistics.
The spectrum therefore contains hidden structure that is obscured in a global level-spacing analysis but becomes visible once the resonances are grouped according to the differential magnetic moments of the associated molecular states.
This behavior can be understood from the large number of molecular configurations with similar magnetic moments.
States near the center of the accessible distribution can arise from many combinations of microscopic quantum numbers with comparable magnetic dependence, and therefore have many nearby molecular states with which they can mix and repel.
By contrast, states near the lower edge of the magnetic-moment distribution have fewer compatible configurations available and retain a more regular statistical character.

We locate the Feshbach resonances by atom-loss spectroscopy.
We prepare nondegenerate samples of $^{162}$Dy, spin-polarized in the $\ket{-8}$ state, with a temperature of $350~{\rm nK}$ and an atom number of $1$--$2\times10^5$. 
The atoms are brought to a chosen magnetic field and held for $500~{\rm ms}$ before the remaining atom number is measured~\cite{SuppMat}. 
Enhanced losses due to two- and three-body recombination processes reveal 128 resonances between 0 and 30~G. 
Their positions $B_i$ are extracted from the loss features shown in Fig.~\ref{figFFRscan}.
Note that only about one third of these resonances had been previously reported~\cite{Baumann2014,Maier2015b,Lucioni2018,Kao2021,Lecomte2024,Duerbeck2026}. 

We determine the differential magnetic moments of the associated molecular states from their near-threshold dispersion, measured by bound-state spectroscopy using a time-modulated spin-dependent light shift~\cite{Journeaux2026,SuppMat}.
The measured binding energies are fitted locally with
\be
E(B)=\delta\mu_i \ (B-B_i),
\ee
where $\delta\mu_i =  \mu_{{\rm mol}}^i - \mu_{\rm en}$ is the differential magnetic moment between the molecular state $\mu^i_{{\rm mol}}$ and the entrance-channel atom pair $\mu_{\rm en} \simeq -19.87 \ \mu_B$.
In total, we determine $\delta\mu$ for 83 of the 128 observed resonances.
All reported values of $\delta\mu$ are obtained by tracking the variation of $E(B)/h$ on (at least) the frequency range $\Delta=[-200,0]~\rm{kHz}$ \cite{SuppMat}.

The resulting distribution of differential magnetic moments provides an experimental characterization of the molecular states responsible for the low-field Feshbach spectrum.
We find a mean value $\overline{\delta\mu}=10.7 \ \mu_B$.
This value is close to the prediction of Ref.~\cite{Mccann2021}, which models  weakly bound Dy$_2$ states in the zero-field $J=16$, $M=-16$ manifold connected to the spin-polarized entrance channel.
In that model, the fully coupled distribution of molecular magnetic moments gives, after subtracting the entrance-channel magnetic moment, $\overline{\delta\mu}^{\rm th}\simeq12.6 \ \mu_B$.

\begin{figure}[t!]
    \centering
    \includegraphics[width=\columnwidth]{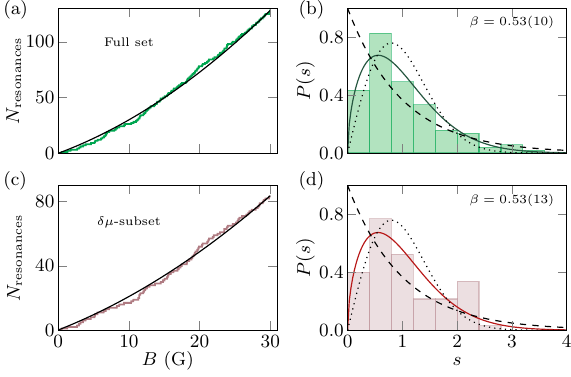}
\caption{
Brody analysis of the $^{162}$Dy Feshbach spectrum. 
Statistical analysis of the resonance spectrum. (a) Unfolding of the complete set of 128 resonances used in the analysis. (b) 
Corresponding nearest-neighbor spacing distribution $P(s)$. The binned histograms are shown only for visualization. The solid line shows the optimized maximum-likelihood Brody distribution, while the Poisson and Wigner--Dyson limits are shown as dashed and dotted lines, respectively. Panels (c) and (d) show the same analysis restricted to the subset of 83 resonances for which differential magnetic moments were experimentally determined. Specifically, (c) presents the unfolding procedure and (d) the resulting spacing distribution $P(s)$. The extracted Brody parameters are $\beta=0.53(10)$ and $\beta=0.53(13)$, for the full set and the 83 resonances set, respectively, with one-standard-deviation uncertainties obtained from the covariance matrix of the log-likelihood estimator~\cite{CasellaBerger2002}.
}
    \label{figBrody}
\end{figure}

Before resolving the spectrum according to $\delta\mu$, we first analyze the complete set of 128 resonances. 
This provides a direct comparison with previous studies based only on resonance positions~\cite{Frisch2014,Maier2015,Khlebnikov2019} and establishes a reference for the magnetic-moment-resolved analysis.
Since the resonance density varies with magnetic field, the raw spacings $d_i = B_{i+1}-B_i$ cannot be compared directly across the full spectrum. 
We therefore unfold the spectrum by fitting the accumulated number of resonances $N(B)$ with a second-order polynomial $g(B)$, constrained by $g(B_{\rm min})=1$ and $g(B_{\rm max})=N_{\rm res}$, where here $B_{\rm max}=29.82~{\rm G}$, $B_{\rm min}=0.492~{\rm G}$ and $N_{\rm res}=128$ (see Fig.~\ref{figBrody}). 
The unfolded nearest-neighbor spacings,
\be
s_i=g(B_{i+1})-g(B_i),
\ee
remove the smooth magnetic-field dependence of the resonance density and have unit mean spacing.
We compare the resulting spacing distribution with the Brody distribution,
\be
P_{\beta}(s)=\alpha (\beta+1)s^\beta
e^{-\alpha s^{\beta+1}},
\label{Brodyeq}
\ee
where $\alpha$
fixes the mean spacing to unity~\footnote{ In practice this corresponds to having \be \alpha= \left[ \Gamma\!\left(\frac{\beta+2}{\beta+1}\right) \right]^{\beta+1}\ , \ee  where $\Gamma(x)$ is the gamma function. }.
The Brody parameter $\beta$ interpolates between the Poisson limit, $\beta=0$, expected for uncorrelated levels, and the Wigner-Dyson limit, $\beta=1$, characteristic of level repulsion.
To avoid any dependence on histogram binning, we extract $\beta$ directly from the unfolded spacings by maximizing the log-likelihood
\be
\mathcal L(\beta)=\sum_i \log P_{\beta}(s_i).
\ee
For the set of 128 resonances, this gives $\beta=0.53(10)$~\footnote{For completeness, we also test for a possible smooth magnetic-field dependence of the Brody parameter by writing $\beta_i=\beta_0+\beta_1\bar B_i$, where $\bar B_i=(B_{i+1}+B_i)/2$~\cite{Makrides2018,Maier2015}. The resulting fit gives $\beta_0=0.54(10)$ and $\beta_1=-0.06(37)~{\rm G}^{-1}$, consistent with no measurable field dependence over the investigated range.
}.

Applying the same analysis to the subset of 83 resonances for which $\delta\mu$ is measured gives $\beta=0.53(13)$, in agreement with the complete set.
This shows that the spectroscopically characterized resonances preserve the global statistical properties of the full Feshbach spectrum (see Fig.~\ref{figBrody}).

\begin{figure}[t!]
    \centering
    \includegraphics[width=\columnwidth]{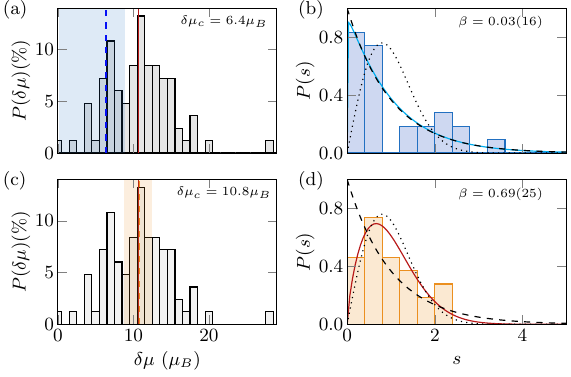}
\caption{
Magnetic-moment-resolved Brody analysis of the $^{162}$Dy Feshbach spectrum. Resonances are selected according to their measured differential magnetic moment $\delta\mu$. Panels (a) and (c) show the measured differential-magnetic-moment distributions, with shaded regions indicating representative selections. The dashed line marks the central value $\delta\mu_c$ of each subset, defined as the mean differential magnetic moment of the selected resonances, while the red line marks the mean of the full distribution, $\overline{\delta\mu}=10.7\ \mu_{\rm B}$. Panels (b) and (d) show the corresponding unfolded nearest-neighbor spacing distributions $P(s)$, with $s_i=g(B_{i+1})-g(B_i)$. The binned histograms are shown only for visualization. Solid lines show Brody distributions evaluated with the values of $\beta$ obtained from unbinned maximum-log-likelihood estimation of the individual spacings, as indicated in the legends. The Poisson (dashed line) and Wigner-Dyson (dotted line) limits are included for comparison. Each selection contains 28 Fano-Feshbach resonances.
}
\label{figBrodyMu}
\end{figure}

We then resolve the statistics according to the measured differential magnetic moment, which constitutes the main result of this work. As shown in Fig.~\ref{figBrodyMu}, we select two representative subsets each containing 28 resonances: a low-moment subset with central differential magnetic moment $\delta\mu_c=6.4\ \mu_{\rm B}$ and a central subset with $\delta\mu_c=10.8\ \mu_{\rm B}$, close to the mean value $\overline{\delta\mu}=10.7\ \mu_{\rm B}$~\footnote{$\delta\mu_c$ is defined as the mean value of $\delta\mu$ over the selected resonances.
}. The two subsets do not overlap with each other. For each case we independently unfold the resonance positions and extract the Brody parameter by maximum-log-likelihood estimation. 
We restrict the magnetic-moment-resolved statistical analysis to $\delta\mu<15.3\,\mu_{\rm B}$, where the spectroscopy provides enough near-threshold points to determine $\delta\mu$ accurately.

For the low differential magnetic moment subset, \emph{i.e.} $\delta \mu_c = 6.4 \ \mu_B$, the Brody parameter is $\beta=0.03(16)$. This value is close to the Poissonian limit, indicating very weak level repulsion.
For the central subset, the spacing distribution displays enhanced level repulsion, with $\beta=0.69(25)$, moving toward the Wigner-Dyson limit.
The chaotic character of the Feshbach spectrum is therefore not uniform across the molecular spectrum, but depends on the magnetic nature of the associated molecular states.

As a complementary diagnostic, we analyze the average gap ratio $\av{r}$, which does not require unfolding and is insensitive to the smooth variation of the resonance density~\cite{Oganesyan2007,Atas2013,Giraud2022}. 
For consecutive resonance spacings in magnetic field, $d_i=B_{i+1}-B_i$, we define
\be
r_i=\min\pr{d_i/d_{i-1},d_{i-1}/d_i}.
\ee
The mean value $\av{r}$ distinguishes the Poisson limit, $\av{r}=2\ln 2-1\simeq0.386$, from the Gaussian orthogonal ensemble (GOE) value, $\av{r}\simeq0.531$, expected for Wigner-Dyson spectral correlations. 
This quantity is complementary to the Brody parameter: $\beta$ is extracted from the nearest-neighbor spacing distribution $P(s)$, whereas $\av{r}$ depends on pairs of consecutive spacings~\cite{SuppMat}.
For the complete set of 128 Feshbach resonances we find $\av{r}=0.49(2)$, while the subset of 83 resonances for which $\delta\mu$ is measured gives $\av{r}=0.47(3)$, consistent with the intermediate level repulsion inferred from the Brody analysis.
Resolving the spectrum by differential magnetic moment confirms our main result: the low differential magnetic moment subset gives $\av{r}=0.34(6)$, close to the Poisson limit, whereas the central subset with $\delta \mu_c = 10.8 \ \mu_B$ gives $\av{r}=0.52(5)$. 
A sliding-window analysis, including the experimental uncertainties on the measured $\delta\mu_i$, confirms that this trend is robust against the choice of window position and size~\cite{SuppMat}.
Although the uncertainties are limited by the finite number of resonances, this unfolding-free analysis supports the conclusion that level repulsion is enhanced near the center of the differential-magnetic-moment distribution.

This behavior admits a natural microscopic interpretation. 
The differential magnetic moment reflects the magnetic properties of a molecular state and thereby constrains which molecular configurations can contribute to it. 
States near the lower edge of the measured $\delta\mu$ distribution remain close to the spin-polarized entrance channel composed of two atoms in stretched spin states, and therefore have only few compatible spin configurations available for mixing.
This scarcity of coupling partners naturally leads to weak level repulsion and statistics close to the Poisson limit.
States near the center of the distribution, by contrast, can be realized by many combinations of spin and orbital angular momentum, and are therefore embedded in a dense network of energetically close molecular states. Anisotropic interactions within this network generate strong mixing and enhanced level repulsion, driving the statistics toward the Wigner-Dyson regime~\cite{Gubin2012,Batistic2013,Mccann2021,Kosicki2020}. This picture is consistent with the broader observation that global spectral statistics can conceal statistically distinct components, which become apparent only upon resolving an appropriate conserved or approximate quantum number~\cite{Giraud2022}.

Our approach treats $\delta\mu$ as a spectroscopic marker of molecular-state character, and rests on the assumption that the measured magnetic-moment distribution faithfully represents the underlying molecular spectrum, i.e., that no relevant class of states is missing from the two subsets analyzed. One might worry that the $\delta\mu$ distribution of Fig.\,\ref{figBrodyMu} is centered at $\overline{\delta\mu} = 10.7\,\mu_{\rm B}$, substantially below the midpoint value of $20\,\mu_{\rm B}$
that one would expect if the molecular moments $\mu_{\rm mol}$ were uniformly distributed over $[-20\,\mu_{\rm B}, +20\,\mu_{\rm B}]$. However, as noted in Ref.~\cite{Mccann2021}, this shift is in fact predicted by a simple model that accounts for the structure of the entrance channel:  Since this channel consists of an atom pair in the spin state $\ket{-8,\,-8}$ with orbital angular momentum $L = 0$, all states coupled to it must satisfy $M = -16$, where $M$ is the projection quantum number of the total angular momentum. Reaching a magnetic moment near the upper end of the range, \emph{i.e.} $\mu_{\rm mol} \sim +20\,\mu_{\rm B}$, while satisfying this constraint requires states with very large orbital angular momentum, typically $L \sim 32$. For such states, the large centrifugal barrier increases the vibrational level spacing and thereby substantially reduces the density of states at the energy of the entrance channel, making these configurations strongly underrepresented in the observed spectrum.
In addition to this physical argument, we cannot rule out state-dependent detection effects that would further deplete the upper end of the measured $\delta\mu$ distribution. 
In the present case, resonances associated with large differential magnetic moments may be narrower in magnetic field and therefore harder to resolve with the finite magnetic-field step size used in the scan.

In conclusion, our measurements demonstrate that the differential magnetic moment transforms the Feshbach spectrum from a single, apparently featureless statistical ensemble into a partially state-resolved object. Grouping resonances according to the magnetic moment of the associated molecular states reveals that an apparently intermediate level-spacing distribution in fact separates into subsets with markedly different degrees of level repulsion, a structure that would remain invisible otherwise. This establishes eigenstate-sensitive spectroscopy as a practical route to uncover hidden statistical structure in chaotic lanthanide Feshbach spectra. 

More broadly, our results suggest that the coexistence of regular and chaotic components is not a peculiarity of a single spectrum, but a generic feature of molecular spectra governed by anisotropic interactions. 
In such systems, the conserved total angular-momentum projection $M$ constrains the accessible molecular configurations, while the magnetic moment provides an experimentally accessible classifier of their character. 
A direct test of this picture would be to repeat the experiment in entrance channels with different $M$. 
Changing $M$ should shift the center of the $\delta\mu$ distribution and modify the balance between regular and chaotic sectors, providing a tunable handle on the statistical structure of the spectrum.

\vspace{0.5cm}
\begin{acknowledgments}

 We are grateful to the members of the Bose--Einstein condensate team at LKB for helpful discussions. This research was funded, in part, by the Agence Nationale de la Recherche (ANR) under projects ANR-20-CE30-0024 and ANR-24-CE30-7961. This work was also supported by the Région Île-de-France within the framework of DIM QuanTiP, São Paulo Resarch Foundation (FAPESP) under the grant 2025/19167-0, and by the ``Fondation CFM pour la Recherche''. For the purpose of open access, the author has applied a CC-BY public copyright licence to any Author Accepted Manuscript (AAM) version arising from this submission. 
\end{acknowledgments}

\bibliography{bib.bib}

\begin{thebibliography}{58}%
\makeatletter
\providecommand \@ifxundefined [1]{%
 \@ifx{#1\undefined}
}%
\providecommand \@ifnum [1]{%
 \ifnum #1\expandafter \@firstoftwo
 \else \expandafter \@secondoftwo
 \fi
}%
\providecommand \@ifx [1]{%
 \ifx #1\expandafter \@firstoftwo
 \else \expandafter \@secondoftwo
 \fi
}%
\providecommand \natexlab [1]{#1}%
\providecommand \enquote  [1]{``#1''}%
\providecommand \bibnamefont  [1]{#1}%
\providecommand \bibfnamefont [1]{#1}%
\providecommand \citenamefont [1]{#1}%
\providecommand \href@noop [0]{\@secondoftwo}%
\providecommand \href [0]{\begingroup \@sanitize@url \@href}%
\providecommand \@href[1]{\@@startlink{#1}\@@href}%
\providecommand \@@href[1]{\endgroup#1\@@endlink}%
\providecommand \@sanitize@url [0]{\catcode `\\12\catcode `\$12\catcode
  `\&12\catcode `\#12\catcode `\^12\catcode `\_12\catcode `\%12\relax}%
\providecommand \@@startlink[1]{}%
\providecommand \@@endlink[0]{}%
\providecommand \url  [0]{\begingroup\@sanitize@url \@url }%
\providecommand \@url [1]{\endgroup\@href {#1}{\urlprefix }}%
\providecommand \urlprefix  [0]{URL }%
\providecommand \Eprint [0]{\href }%
\providecommand \doibase [0]{https://doi.org/}%
\providecommand \selectlanguage [0]{\@gobble}%
\providecommand \bibinfo  [0]{\@secondoftwo}%
\providecommand \bibfield  [0]{\@secondoftwo}%
\providecommand \translation [1]{[#1]}%
\providecommand \BibitemOpen [0]{}%
\providecommand \bibitemStop [0]{}%
\providecommand \bibitemNoStop [0]{.\EOS\space}%
\providecommand \EOS [0]{\spacefactor3000\relax}%
\providecommand \BibitemShut  [1]{\csname bibitem#1\endcsname}%
\let\auto@bib@innerbib\@empty
\bibitem [{\citenamefont {Baier}\ \emph {et~al.}(2016)\citenamefont {Baier},
  \citenamefont {Mark}, \citenamefont {Petter}, \citenamefont {Aikawa},
  \citenamefont {Chomaz}, \citenamefont {Cai}, \citenamefont {Baranov},
  \citenamefont {Zoller},\ and\ \citenamefont {Ferlaino}}]{Baier2016}%
  \BibitemOpen
  \bibfield  {author} {\bibinfo {author} {\bibfnamefont {S.}~\bibnamefont
  {Baier}}, \bibinfo {author} {\bibfnamefont {M.~J.}\ \bibnamefont {Mark}},
  \bibinfo {author} {\bibfnamefont {D.}~\bibnamefont {Petter}}, \bibinfo
  {author} {\bibfnamefont {K.}~\bibnamefont {Aikawa}}, \bibinfo {author}
  {\bibfnamefont {L.}~\bibnamefont {Chomaz}}, \bibinfo {author} {\bibfnamefont
  {Z.}~\bibnamefont {Cai}}, \bibinfo {author} {\bibfnamefont {M.}~\bibnamefont
  {Baranov}}, \bibinfo {author} {\bibfnamefont {P.}~\bibnamefont {Zoller}},\
  and\ \bibinfo {author} {\bibfnamefont {F.}~\bibnamefont {Ferlaino}},\
  }\bibfield  {title} {\bibinfo {title} {{Extended Bose-Hubbard models with
  ultracold magnetic atoms}},\ }\href@noop {} {\bibfield  {journal} {\bibinfo
  {journal} {Science}\ }\textbf {\bibinfo {volume} {352}},\ \bibinfo {pages}
  {201} (\bibinfo {year} {2016})}\BibitemShut {NoStop}%
\bibitem [{\citenamefont {Su}\ \emph {et~al.}(2023)\citenamefont {Su},
  \citenamefont {Douglas}, \citenamefont {Szurek}, \citenamefont {Groth},
  \citenamefont {Ozturk}, \citenamefont {Krahn}, \citenamefont {H{\'e}bert},
  \citenamefont {Phelps}, \citenamefont {Ebadi}, \citenamefont {Dickerson},
  \citenamefont {Ferlaino}, \citenamefont {Markovi\'c},\ and\ \citenamefont
  {Greiner}}]{Su2023}%
  \BibitemOpen
  \bibfield  {author} {\bibinfo {author} {\bibfnamefont {L.}~\bibnamefont
  {Su}}, \bibinfo {author} {\bibfnamefont {A.}~\bibnamefont {Douglas}},
  \bibinfo {author} {\bibfnamefont {M.}~\bibnamefont {Szurek}}, \bibinfo
  {author} {\bibfnamefont {R.}~\bibnamefont {Groth}}, \bibinfo {author}
  {\bibfnamefont {S.~F.}\ \bibnamefont {Ozturk}}, \bibinfo {author}
  {\bibfnamefont {A.}~\bibnamefont {Krahn}}, \bibinfo {author} {\bibfnamefont
  {A.~H.}\ \bibnamefont {H{\'e}bert}}, \bibinfo {author} {\bibfnamefont
  {G.~A.}\ \bibnamefont {Phelps}}, \bibinfo {author} {\bibfnamefont
  {S.}~\bibnamefont {Ebadi}}, \bibinfo {author} {\bibfnamefont
  {S.}~\bibnamefont {Dickerson}}, \bibinfo {author} {\bibfnamefont
  {F.}~\bibnamefont {Ferlaino}}, \bibinfo {author} {\bibfnamefont
  {O.}~\bibnamefont {Markovi\'c}},\ and\ \bibinfo {author} {\bibfnamefont
  {M.}~\bibnamefont {Greiner}},\ }\bibfield  {title} {\bibinfo {title}
  {{Dipolar quantum solids emerging in a Hubbard quantum simulator}},\
  }\href@noop {} {\bibfield  {journal} {\bibinfo  {journal} {Nature}\ }\textbf
  {\bibinfo {volume} {622}},\ \bibinfo {pages} {724} (\bibinfo {year}
  {2023})}\BibitemShut {NoStop}%
\bibitem [{\citenamefont {Chomaz}\ \emph {et~al.}(2022)\citenamefont {Chomaz},
  \citenamefont {Ferrier-Barbut}, \citenamefont {Ferlaino}, \citenamefont
  {Laburthe-Tolra}, \citenamefont {Lev},\ and\ \citenamefont
  {Pfau}}]{Chomaz2023}%
  \BibitemOpen
  \bibfield  {author} {\bibinfo {author} {\bibfnamefont {L.}~\bibnamefont
  {Chomaz}}, \bibinfo {author} {\bibfnamefont {I.}~\bibnamefont
  {Ferrier-Barbut}}, \bibinfo {author} {\bibfnamefont {F.}~\bibnamefont
  {Ferlaino}}, \bibinfo {author} {\bibfnamefont {B.}~\bibnamefont
  {Laburthe-Tolra}}, \bibinfo {author} {\bibfnamefont {B.~L.}\ \bibnamefont
  {Lev}},\ and\ \bibinfo {author} {\bibfnamefont {T.}~\bibnamefont {Pfau}},\
  }\bibfield  {title} {\bibinfo {title} {Dipolar physics: a review of
  experiments with magnetic quantum gases},\ }\href
  {https://doi.org/10.1088/1361-6633/aca814} {\bibfield  {journal} {\bibinfo
  {journal} {Rep. Prog. Phys.}\ }\textbf {\bibinfo {volume} {86}},\ \bibinfo
  {pages} {026401} (\bibinfo {year} {2022})}\BibitemShut {NoStop}%
\bibitem [{\citenamefont {Biagioni}\ \emph {et~al.}(2024)\citenamefont
  {Biagioni}, \citenamefont {Antolini}, \citenamefont {Donelli}, \citenamefont
  {Pezz{\`e}}, \citenamefont {Smerzi}, \citenamefont {Fattori}, \citenamefont
  {Fioretti}, \citenamefont {Gabbanini}, \citenamefont {Inguscio},
  \citenamefont {Tanzi},\ and\ \citenamefont {Modugno}}]{Biagioni2024}%
  \BibitemOpen
  \bibfield  {author} {\bibinfo {author} {\bibfnamefont {G.}~\bibnamefont
  {Biagioni}}, \bibinfo {author} {\bibfnamefont {N.}~\bibnamefont {Antolini}},
  \bibinfo {author} {\bibfnamefont {B.}~\bibnamefont {Donelli}}, \bibinfo
  {author} {\bibfnamefont {L.}~\bibnamefont {Pezz{\`e}}}, \bibinfo {author}
  {\bibfnamefont {A.}~\bibnamefont {Smerzi}}, \bibinfo {author} {\bibfnamefont
  {M.}~\bibnamefont {Fattori}}, \bibinfo {author} {\bibfnamefont
  {A.}~\bibnamefont {Fioretti}}, \bibinfo {author} {\bibfnamefont
  {C.}~\bibnamefont {Gabbanini}}, \bibinfo {author} {\bibfnamefont
  {M.}~\bibnamefont {Inguscio}}, \bibinfo {author} {\bibfnamefont
  {L.}~\bibnamefont {Tanzi}},\ and\ \bibinfo {author} {\bibfnamefont
  {G.}~\bibnamefont {Modugno}},\ }\bibfield  {title} {\bibinfo {title}
  {{Measurement of the superfluid fraction of a supersolid by Josephson
  effect}},\ }\href {https://doi.org/10.1038/s41586-024-07361-9} {\bibfield
  {journal} {\bibinfo  {journal} {Nature}\ }\textbf {\bibinfo {volume} {629}},\
  \bibinfo {pages} {773} (\bibinfo {year} {2024})}\BibitemShut {NoStop}%
\bibitem [{\citenamefont {Lecomte}\ \emph {et~al.}(2025)\citenamefont
  {Lecomte}, \citenamefont {Journeaux}, \citenamefont {Veschambre},
  \citenamefont {Dalibard},\ and\ \citenamefont {Lopes}}]{Lecomte2025}%
  \BibitemOpen
  \bibfield  {author} {\bibinfo {author} {\bibfnamefont {M.}~\bibnamefont
  {Lecomte}}, \bibinfo {author} {\bibfnamefont {A.}~\bibnamefont {Journeaux}},
  \bibinfo {author} {\bibfnamefont {J.}~\bibnamefont {Veschambre}}, \bibinfo
  {author} {\bibfnamefont {J.}~\bibnamefont {Dalibard}},\ and\ \bibinfo
  {author} {\bibfnamefont {R.}~\bibnamefont {Lopes}},\ }\bibfield  {title}
  {\bibinfo {title} {{Production and Stabilization of a Spin Mixture of
  Ultracold Dipolar Bose Gases}},\ }\href
  {https://doi.org/10.1103/PhysRevLett.134.013402} {\bibfield  {journal}
  {\bibinfo  {journal} {Phys. Rev. Lett.}\ }\textbf {\bibinfo {volume} {134}},\
  \bibinfo {pages} {013402} (\bibinfo {year} {2025})}\BibitemShut {NoStop}%
\bibitem [{\citenamefont {Lafforgue}\ \emph {et~al.}(2025)\citenamefont
  {Lafforgue}, \citenamefont {Mehta}, \citenamefont {Houwman}, \citenamefont
  {Claude}, \citenamefont {Rittenhouse}, \citenamefont {Ferlaino},\ and\
  \citenamefont {Mark}}]{Lafforgue2025}%
  \BibitemOpen
  \bibfield  {author} {\bibinfo {author} {\bibfnamefont {L.}~\bibnamefont
  {Lafforgue}}, \bibinfo {author} {\bibfnamefont {N.}~\bibnamefont {Mehta}},
  \bibinfo {author} {\bibfnamefont {J.}~\bibnamefont {Houwman}}, \bibinfo
  {author} {\bibfnamefont {F.}~\bibnamefont {Claude}}, \bibinfo {author}
  {\bibfnamefont {S.}~\bibnamefont {Rittenhouse}}, \bibinfo {author}
  {\bibfnamefont {F.}~\bibnamefont {Ferlaino}},\ and\ \bibinfo {author}
  {\bibfnamefont {M.}~\bibnamefont {Mark}},\ }\bibfield  {title} {\bibinfo
  {title} {{Observation of Fano-suppression in scattering resonances of bosonic
  erbium atoms}},\ }\href@noop {} {\bibfield  {journal} {\bibinfo  {journal}
  {arXiv:2512.17556}\ } (\bibinfo {year} {2025})}\BibitemShut {NoStop}%
\bibitem [{\citenamefont {Gawryluk}\ \emph {et~al.}(2007)\citenamefont
  {Gawryluk}, \citenamefont {Brewczyk}, \citenamefont {Bongs},\ and\
  \citenamefont {Gajda}}]{Gawryluk07}%
  \BibitemOpen
  \bibfield  {author} {\bibinfo {author} {\bibfnamefont {K.}~\bibnamefont
  {Gawryluk}}, \bibinfo {author} {\bibfnamefont {M.}~\bibnamefont {Brewczyk}},
  \bibinfo {author} {\bibfnamefont {K.}~\bibnamefont {Bongs}},\ and\ \bibinfo
  {author} {\bibfnamefont {M.}~\bibnamefont {Gajda}},\ }\bibfield  {title}
  {\bibinfo {title} {{Resonant Einstein--de Haas Effect in a Rubidium
  Condensate}},\ }\href {https://doi.org/10.1103/PhysRevLett.99.130401}
  {\bibfield  {journal} {\bibinfo  {journal} {Phys. Rev. Lett.}\ }\textbf
  {\bibinfo {volume} {99}},\ \bibinfo {pages} {130401} (\bibinfo {year}
  {2007})}\BibitemShut {NoStop}%
\bibitem [{\citenamefont {\ifmmode~\acute{S}\else \'{S}\fi{}wis\l{}ocki}\ \emph
  {et~al.}(2011)\citenamefont {\ifmmode~\acute{S}\else \'{S}\fi{}wis\l{}ocki},
  \citenamefont {Sowi\ifmmode~\acute{n}\else \'{n}\fi{}ski}, \citenamefont
  {Pietraszewicz}, \citenamefont {Brewczyk}, \citenamefont {Lewenstein},
  \citenamefont {Zakrzewski},\ and\ \citenamefont {Gajda}}]{Swi2011}%
  \BibitemOpen
  \bibfield  {author} {\bibinfo {author} {\bibfnamefont {T.}~\bibnamefont
  {\ifmmode~\acute{S}\else \'{S}\fi{}wis\l{}ocki}}, \bibinfo {author}
  {\bibfnamefont {T.}~\bibnamefont {Sowi\ifmmode~\acute{n}\else
  \'{n}\fi{}ski}}, \bibinfo {author} {\bibfnamefont {J.}~\bibnamefont
  {Pietraszewicz}}, \bibinfo {author} {\bibfnamefont {M.}~\bibnamefont
  {Brewczyk}}, \bibinfo {author} {\bibfnamefont {M.}~\bibnamefont
  {Lewenstein}}, \bibinfo {author} {\bibfnamefont {J.}~\bibnamefont
  {Zakrzewski}},\ and\ \bibinfo {author} {\bibfnamefont {M.}~\bibnamefont
  {Gajda}},\ }\bibfield  {title} {\bibinfo {title} {{Tunable dipolar resonances
  and Einstein-de Haas effect in a $^{87}\mathrm{Rb}$-atom condensate}},\
  }\href {https://doi.org/10.1103/PhysRevA.83.063617} {\bibfield  {journal}
  {\bibinfo  {journal} {Phys. Rev. A}\ }\textbf {\bibinfo {volume} {83}},\
  \bibinfo {pages} {063617} (\bibinfo {year} {2011})}\BibitemShut {NoStop}%
\bibitem [{\citenamefont {Matsui}\ \emph {et~al.}(2026)\citenamefont {Matsui},
  \citenamefont {Miyazawa}, \citenamefont {Goto}, \citenamefont {Nakano},
  \citenamefont {Kawaguchi}, \citenamefont {Ueda},\ and\ \citenamefont
  {Kozuma}}]{Matsui2026}%
  \BibitemOpen
  \bibfield  {author} {\bibinfo {author} {\bibfnamefont {H.}~\bibnamefont
  {Matsui}}, \bibinfo {author} {\bibfnamefont {Y.}~\bibnamefont {Miyazawa}},
  \bibinfo {author} {\bibfnamefont {R.}~\bibnamefont {Goto}}, \bibinfo {author}
  {\bibfnamefont {C.}~\bibnamefont {Nakano}}, \bibinfo {author} {\bibfnamefont
  {Y.}~\bibnamefont {Kawaguchi}}, \bibinfo {author} {\bibfnamefont
  {M.}~\bibnamefont {Ueda}},\ and\ \bibinfo {author} {\bibfnamefont
  {M.}~\bibnamefont {Kozuma}},\ }\bibfield  {title} {\bibinfo {title}
  {{Observation of the Einstein--de Haas effect in a Bose--Einstein
  condensate}},\ }\href@noop {} {\bibfield  {journal} {\bibinfo  {journal}
  {Science}\ }\textbf {\bibinfo {volume} {391}},\ \bibinfo {pages} {384}
  (\bibinfo {year} {2026})}\BibitemShut {NoStop}%
\bibitem [{\citenamefont {Baumann}\ \emph {et~al.}(2014)\citenamefont
  {Baumann}, \citenamefont {Burdick}, \citenamefont {Lu},\ and\ \citenamefont
  {Lev}}]{Baumann2014}%
  \BibitemOpen
  \bibfield  {author} {\bibinfo {author} {\bibfnamefont {K.}~\bibnamefont
  {Baumann}}, \bibinfo {author} {\bibfnamefont {N.~Q.}\ \bibnamefont
  {Burdick}}, \bibinfo {author} {\bibfnamefont {M.}~\bibnamefont {Lu}},\ and\
  \bibinfo {author} {\bibfnamefont {B.~L.}\ \bibnamefont {Lev}},\ }\bibfield
  {title} {\bibinfo {title} {{Observation of low-field Fano-Feshbach resonances
  in ultracold gases of dysprosium}},\ }\href
  {https://doi.org/10.1103/PhysRevA.89.020701} {\bibfield  {journal} {\bibinfo
  {journal} {Phys. Rev. A}\ }\textbf {\bibinfo {volume} {89}},\ \bibinfo
  {pages} {020701} (\bibinfo {year} {2014})}\BibitemShut {NoStop}%
\bibitem [{\citenamefont {Frisch}\ \emph {et~al.}(2015)\citenamefont {Frisch},
  \citenamefont {Mark}, \citenamefont {Aikawa}, \citenamefont {Baier},
  \citenamefont {Grimm}, \citenamefont {Petrov}, \citenamefont {Kotochigova},
  \citenamefont {Qu{\'e}m{\'e}ner}, \citenamefont {Lepers}, \citenamefont
  {Dulieu},\ and\ \citenamefont {Ferlaino}}]{Frisch2015}%
  \BibitemOpen
  \bibfield  {author} {\bibinfo {author} {\bibfnamefont {A.}~\bibnamefont
  {Frisch}}, \bibinfo {author} {\bibfnamefont {M.}~\bibnamefont {Mark}},
  \bibinfo {author} {\bibfnamefont {K.}~\bibnamefont {Aikawa}}, \bibinfo
  {author} {\bibfnamefont {S.}~\bibnamefont {Baier}}, \bibinfo {author}
  {\bibfnamefont {R.}~\bibnamefont {Grimm}}, \bibinfo {author} {\bibfnamefont
  {A.}~\bibnamefont {Petrov}}, \bibinfo {author} {\bibfnamefont
  {S.}~\bibnamefont {Kotochigova}}, \bibinfo {author} {\bibfnamefont
  {G.}~\bibnamefont {Qu{\'e}m{\'e}ner}}, \bibinfo {author} {\bibfnamefont
  {M.}~\bibnamefont {Lepers}}, \bibinfo {author} {\bibfnamefont
  {O.}~\bibnamefont {Dulieu}},\ and\ \bibinfo {author} {\bibfnamefont
  {F.}~\bibnamefont {Ferlaino}},\ }\bibfield  {title} {\bibinfo {title}
  {{Ultracold dipolar molecules composed of strongly magnetic atoms}},\
  }\href@noop {} {\bibfield  {journal} {\bibinfo  {journal} {Phys. Rev. Lett.}\
  }\textbf {\bibinfo {volume} {115}},\ \bibinfo {pages} {203201} (\bibinfo
  {year} {2015})}\BibitemShut {NoStop}%
\bibitem [{\citenamefont {Maier}\ \emph
  {et~al.}(2015{\natexlab{a}})\citenamefont {Maier}, \citenamefont {Kadau},
  \citenamefont {Schmitt}, \citenamefont {Wenzel}, \citenamefont
  {Ferrier-Barbut}, \citenamefont {Pfau}, \citenamefont {Frisch}, \citenamefont
  {Baier}, \citenamefont {Aikawa}, \citenamefont {Chomaz}, \citenamefont
  {Mark}, \citenamefont {Ferlaino}, \citenamefont {Makrides}, \citenamefont
  {Tiesinga}, \citenamefont {Petrov},\ and\ \citenamefont
  {Kotochigova}}]{Maier2015}%
  \BibitemOpen
  \bibfield  {author} {\bibinfo {author} {\bibfnamefont {T.}~\bibnamefont
  {Maier}}, \bibinfo {author} {\bibfnamefont {H.}~\bibnamefont {Kadau}},
  \bibinfo {author} {\bibfnamefont {M.}~\bibnamefont {Schmitt}}, \bibinfo
  {author} {\bibfnamefont {M.}~\bibnamefont {Wenzel}}, \bibinfo {author}
  {\bibfnamefont {I.}~\bibnamefont {Ferrier-Barbut}}, \bibinfo {author}
  {\bibfnamefont {T.}~\bibnamefont {Pfau}}, \bibinfo {author} {\bibfnamefont
  {A.}~\bibnamefont {Frisch}}, \bibinfo {author} {\bibfnamefont
  {S.}~\bibnamefont {Baier}}, \bibinfo {author} {\bibfnamefont
  {K.}~\bibnamefont {Aikawa}}, \bibinfo {author} {\bibfnamefont
  {L.}~\bibnamefont {Chomaz}}, \bibinfo {author} {\bibfnamefont {M.~J.}\
  \bibnamefont {Mark}}, \bibinfo {author} {\bibfnamefont {F.}~\bibnamefont
  {Ferlaino}}, \bibinfo {author} {\bibfnamefont {C.}~\bibnamefont {Makrides}},
  \bibinfo {author} {\bibfnamefont {E.}~\bibnamefont {Tiesinga}}, \bibinfo
  {author} {\bibfnamefont {A.}~\bibnamefont {Petrov}},\ and\ \bibinfo {author}
  {\bibfnamefont {S.}~\bibnamefont {Kotochigova}},\ }\bibfield  {title}
  {\bibinfo {title} {{Emergence of Chaotic Scattering in Ultracold Er and
  Dy}},\ }\href {https://doi.org/10.1103/PhysRevX.5.041029} {\bibfield
  {journal} {\bibinfo  {journal} {Phys. Rev. X}\ }\textbf {\bibinfo {volume}
  {5}},\ \bibinfo {pages} {041029} (\bibinfo {year}
  {2015}{\natexlab{a}})}\BibitemShut {NoStop}%
\bibitem [{\citenamefont {Khlebnikov}\ \emph {et~al.}(2019)\citenamefont
  {Khlebnikov}, \citenamefont {Pershin}, \citenamefont {Tsyganok},
  \citenamefont {Davletov}, \citenamefont {Cojocaru}, \citenamefont {Fedorova},
  \citenamefont {Buchachenko},\ and\ \citenamefont {Akimov}}]{Khlebnikov2019}%
  \BibitemOpen
  \bibfield  {author} {\bibinfo {author} {\bibfnamefont {V.~A.}\ \bibnamefont
  {Khlebnikov}}, \bibinfo {author} {\bibfnamefont {D.~A.}\ \bibnamefont
  {Pershin}}, \bibinfo {author} {\bibfnamefont {V.~V.}\ \bibnamefont
  {Tsyganok}}, \bibinfo {author} {\bibfnamefont {E.~T.}\ \bibnamefont
  {Davletov}}, \bibinfo {author} {\bibfnamefont {I.~S.}\ \bibnamefont
  {Cojocaru}}, \bibinfo {author} {\bibfnamefont {E.~S.}\ \bibnamefont
  {Fedorova}}, \bibinfo {author} {\bibfnamefont {A.~A.}\ \bibnamefont
  {Buchachenko}},\ and\ \bibinfo {author} {\bibfnamefont {A.~V.}\ \bibnamefont
  {Akimov}},\ }\bibfield  {title} {\bibinfo {title} {{Random to Chaotic
  Statistic Transformation in Low-Field Fano-Feshbach Resonances of Cold
  Thulium Atoms}},\ }\href {https://doi.org/10.1103/PhysRevLett.123.213402}
  {\bibfield  {journal} {\bibinfo  {journal} {Phys. Rev. Lett.}\ }\textbf
  {\bibinfo {volume} {123}},\ \bibinfo {pages} {213402} (\bibinfo {year}
  {2019})}\BibitemShut {NoStop}%
\bibitem [{\citenamefont {Frisch}\ \emph {et~al.}(2014)\citenamefont {Frisch},
  \citenamefont {Mark}, \citenamefont {Aikawa}, \citenamefont {Ferlaino},
  \citenamefont {Bohn}, \citenamefont {Makrides}, \citenamefont {Petrov},\ and\
  \citenamefont {Kotochigova}}]{Frisch2014}%
  \BibitemOpen
  \bibfield  {author} {\bibinfo {author} {\bibfnamefont {A.}~\bibnamefont
  {Frisch}}, \bibinfo {author} {\bibfnamefont {M.}~\bibnamefont {Mark}},
  \bibinfo {author} {\bibfnamefont {K.}~\bibnamefont {Aikawa}}, \bibinfo
  {author} {\bibfnamefont {F.}~\bibnamefont {Ferlaino}}, \bibinfo {author}
  {\bibfnamefont {J.~L.}\ \bibnamefont {Bohn}}, \bibinfo {author}
  {\bibfnamefont {C.}~\bibnamefont {Makrides}}, \bibinfo {author}
  {\bibfnamefont {A.}~\bibnamefont {Petrov}},\ and\ \bibinfo {author}
  {\bibfnamefont {S.}~\bibnamefont {Kotochigova}},\ }\bibfield  {title}
  {\bibinfo {title} {{Quantum chaos in ultracold collisions of gas-phase erbium
  atoms}},\ }\href {https://doi.org/10.1038/nature13137} {\bibfield  {journal}
  {\bibinfo  {journal} {Nature}\ }\textbf {\bibinfo {volume} {507}},\ \bibinfo
  {pages} {475} (\bibinfo {year} {2014})}\BibitemShut {NoStop}%
\bibitem [{\citenamefont {Maier}\ \emph
  {et~al.}(2015{\natexlab{b}})\citenamefont {Maier}, \citenamefont
  {Ferrier-Barbut}, \citenamefont {Kadau}, \citenamefont {Schmitt},
  \citenamefont {Wenzel}, \citenamefont {Wink}, \citenamefont {Pfau},
  \citenamefont {Jachymski},\ and\ \citenamefont {Julienne}}]{Maier2015b}%
  \BibitemOpen
  \bibfield  {author} {\bibinfo {author} {\bibfnamefont {T.}~\bibnamefont
  {Maier}}, \bibinfo {author} {\bibfnamefont {I.}~\bibnamefont
  {Ferrier-Barbut}}, \bibinfo {author} {\bibfnamefont {H.}~\bibnamefont
  {Kadau}}, \bibinfo {author} {\bibfnamefont {M.}~\bibnamefont {Schmitt}},
  \bibinfo {author} {\bibfnamefont {M.}~\bibnamefont {Wenzel}}, \bibinfo
  {author} {\bibfnamefont {C.}~\bibnamefont {Wink}}, \bibinfo {author}
  {\bibfnamefont {T.}~\bibnamefont {Pfau}}, \bibinfo {author} {\bibfnamefont
  {K.}~\bibnamefont {Jachymski}},\ and\ \bibinfo {author} {\bibfnamefont
  {P.~S.}\ \bibnamefont {Julienne}},\ }\bibfield  {title} {\bibinfo {title}
  {{Broad universal Feshbach resonances in the chaotic spectrum of dysprosium
  atoms}},\ }\href {https://doi.org/10.1103/PhysRevA.92.060702} {\bibfield
  {journal} {\bibinfo  {journal} {Phys. Rev. A}\ }\textbf {\bibinfo {volume}
  {92}},\ \bibinfo {pages} {060702(R)} (\bibinfo {year}
  {2015}{\natexlab{b}})}\BibitemShut {NoStop}%
\bibitem [{\citenamefont {Kotochigova}\ and\ \citenamefont
  {Petrov}(2011)}]{Kotochigova2011}%
  \BibitemOpen
  \bibfield  {author} {\bibinfo {author} {\bibfnamefont {S.}~\bibnamefont
  {Kotochigova}}\ and\ \bibinfo {author} {\bibfnamefont {A.}~\bibnamefont
  {Petrov}},\ }\bibfield  {title} {\bibinfo {title} {{Anisotropy in the
  interaction of ultracold dysprosium}},\ }\href
  {https://dx.doi.org/10.1039/c1cp21175g} {\bibfield  {journal} {\bibinfo
  {journal} {Phys. Chem. Chem. Phys.}\ }\textbf {\bibinfo {volume} {13}},\
  \bibinfo {pages} {19165} (\bibinfo {year} {2011})}\BibitemShut {NoStop}%
\bibitem [{\citenamefont {Petrov}\ \emph {et~al.}(2012)\citenamefont {Petrov},
  \citenamefont {Tiesinga},\ and\ \citenamefont {Kotochigova}}]{Petrov2012}%
  \BibitemOpen
  \bibfield  {author} {\bibinfo {author} {\bibfnamefont {A.}~\bibnamefont
  {Petrov}}, \bibinfo {author} {\bibfnamefont {E.}~\bibnamefont {Tiesinga}},\
  and\ \bibinfo {author} {\bibfnamefont {S.}~\bibnamefont {Kotochigova}},\
  }\bibfield  {title} {\bibinfo {title} {{Anisotropy-Induced Feshbach
  Resonances in a Quantum Dipolar Gas of Highly Magnetic Atoms}},\ }\href
  {https://doi.org/10.1103/PhysRevLett.109.103002} {\bibfield  {journal}
  {\bibinfo  {journal} {Phys. Rev. Lett.}\ }\textbf {\bibinfo {volume} {109}},\
  \bibinfo {pages} {103002} (\bibinfo {year} {2012})}\BibitemShut {NoStop}%
\bibitem [{\citenamefont {Kotochigova}(2014)}]{Kotochigova2014}%
  \BibitemOpen
  \bibfield  {author} {\bibinfo {author} {\bibfnamefont {S.}~\bibnamefont
  {Kotochigova}},\ }\bibfield  {title} {\bibinfo {title} {{Controlling
  interactions between highly magnetic atoms with Feshbach resonances}},\
  }\href {https://doi.org/10.1088/0034-4885/77/9/093901} {\bibfield  {journal}
  {\bibinfo  {journal} {Rep. Prog. Phys.}\ }\textbf {\bibinfo {volume} {77}},\
  \bibinfo {pages} {093901} (\bibinfo {year} {2014})}\BibitemShut {NoStop}%
\bibitem [{\citenamefont {Makrides}\ \emph {et~al.}(2018)\citenamefont
  {Makrides}, \citenamefont {Li}, \citenamefont {Tiesinga},\ and\ \citenamefont
  {Kotochigova}}]{Makrides2018}%
  \BibitemOpen
  \bibfield  {author} {\bibinfo {author} {\bibfnamefont {C.}~\bibnamefont
  {Makrides}}, \bibinfo {author} {\bibfnamefont {M.}~\bibnamefont {Li}},
  \bibinfo {author} {\bibfnamefont {E.}~\bibnamefont {Tiesinga}},\ and\
  \bibinfo {author} {\bibfnamefont {S.}~\bibnamefont {Kotochigova}},\
  }\bibfield  {title} {\bibinfo {title} {{Fractal universality in
  near-threshold magnetic lanthanide dimers}},\ }\href@noop {} {\bibfield
  {journal} {\bibinfo  {journal} {Science Advances}\ }\textbf {\bibinfo
  {volume} {4}},\ \bibinfo {pages} {eaap8308} (\bibinfo {year}
  {2018})}\BibitemShut {NoStop}%
\bibitem [{\citenamefont {Augustovi{\v{c}}ov{\'a}}\ and\ \citenamefont
  {Bohn}(2018)}]{Augustovivcova2018}%
  \BibitemOpen
  \bibfield  {author} {\bibinfo {author} {\bibfnamefont {L.~D.}\ \bibnamefont
  {Augustovi{\v{c}}ov{\'a}}}\ and\ \bibinfo {author} {\bibfnamefont {J.~L.}\
  \bibnamefont {Bohn}},\ }\bibfield  {title} {\bibinfo {title} {{Manifestation
  of quantum chaos in Fano-Feshbach resonances}},\ }\href@noop {} {\bibfield
  {journal} {\bibinfo  {journal} {Phys. Rev. A}\ }\textbf {\bibinfo {volume}
  {98}},\ \bibinfo {pages} {023419} (\bibinfo {year} {2018})}\BibitemShut
  {NoStop}%
\bibitem [{\citenamefont {McCann}\ \emph {et~al.}(2021)\citenamefont {McCann},
  \citenamefont {Bohn},\ and\ \citenamefont
  {Augustovi{\v{c}}ov{\'a}}}]{Mccann2021}%
  \BibitemOpen
  \bibfield  {author} {\bibinfo {author} {\bibfnamefont {J.}~\bibnamefont
  {McCann}}, \bibinfo {author} {\bibfnamefont {J.~L.}\ \bibnamefont {Bohn}},\
  and\ \bibinfo {author} {\bibfnamefont {L.~D.}\ \bibnamefont
  {Augustovi{\v{c}}ov{\'a}}},\ }\bibfield  {title} {\bibinfo {title} {{Magnetic
  moments of lanthanide van der Waals dimers}},\ }\href@noop {} {\bibfield
  {journal} {\bibinfo  {journal} {Phys. Rev. A}\ }\textbf {\bibinfo {volume}
  {103}},\ \bibinfo {pages} {042812} (\bibinfo {year} {2021})}\BibitemShut
  {NoStop}%
\bibitem [{\citenamefont {Mur-Petit}\ and\ \citenamefont
  {Molina}(2015)}]{Mur2015}%
  \BibitemOpen
  \bibfield  {author} {\bibinfo {author} {\bibfnamefont {J.}~\bibnamefont
  {Mur-Petit}}\ and\ \bibinfo {author} {\bibfnamefont {R.~A.}\ \bibnamefont
  {Molina}},\ }\bibfield  {title} {\bibinfo {title} {{Spectral statistics of
  molecular resonances in erbium isotopes: How chaotic are they?}},\
  }\href@noop {} {\bibfield  {journal} {\bibinfo  {journal} {Phys. Rev. E}\
  }\textbf {\bibinfo {volume} {92}},\ \bibinfo {pages} {042906} (\bibinfo
  {year} {2015})}\BibitemShut {NoStop}%
\bibitem [{\citenamefont {Roy}\ \emph {et~al.}(2017)\citenamefont {Roy},
  \citenamefont {Chakrabarti}, \citenamefont {Chavda}, \citenamefont {Kota},
  \citenamefont {Lekala},\ and\ \citenamefont {Rampho}}]{Roy2017}%
  \BibitemOpen
  \bibfield  {author} {\bibinfo {author} {\bibfnamefont {K.}~\bibnamefont
  {Roy}}, \bibinfo {author} {\bibfnamefont {B.}~\bibnamefont {Chakrabarti}},
  \bibinfo {author} {\bibfnamefont {N.}~\bibnamefont {Chavda}}, \bibinfo
  {author} {\bibfnamefont {V.}~\bibnamefont {Kota}}, \bibinfo {author}
  {\bibfnamefont {M.}~\bibnamefont {Lekala}},\ and\ \bibinfo {author}
  {\bibfnamefont {G.}~\bibnamefont {Rampho}},\ }\bibfield  {title} {\bibinfo
  {title} {{Spectral analysis of molecular resonances in erbium isotopes: Are
  they close to semi-Poisson?}},\ }\href@noop {} {\bibfield  {journal}
  {\bibinfo  {journal} {Europhys. Lett.}\ }\textbf {\bibinfo {volume} {118}},\
  \bibinfo {pages} {46003} (\bibinfo {year} {2017})}\BibitemShut {NoStop}%
\bibitem [{\citenamefont {Casal}\ \emph {et~al.}(2021)\citenamefont {Casal},
  \citenamefont {Mu{\~n}oz},\ and\ \citenamefont {Molina}}]{Casal2021}%
  \BibitemOpen
  \bibfield  {author} {\bibinfo {author} {\bibfnamefont {I.}~\bibnamefont
  {Casal}}, \bibinfo {author} {\bibfnamefont {L.}~\bibnamefont {Mu{\~n}oz}},\
  and\ \bibinfo {author} {\bibfnamefont {R.~A.}\ \bibnamefont {Molina}},\
  }\bibfield  {title} {\bibinfo {title} {{Accuracy and precision of the
  estimation of the number of missing levels in chaotic spectra using
  long-range correlations}},\ }\href@noop {} {\bibfield  {journal} {\bibinfo
  {journal} {The European Physical Journal Plus}\ }\textbf {\bibinfo {volume}
  {136}},\ \bibinfo {pages} {263} (\bibinfo {year} {2021})}\BibitemShut
  {NoStop}%
\bibitem [{\citenamefont {Heller}(1975)}]{Heller75}%
  \BibitemOpen
  \bibfield  {author} {\bibinfo {author} {\bibfnamefont {E.~J.}\ \bibnamefont
  {Heller}},\ }\bibfield  {title} {\bibinfo {title} {{Time‐dependent approach
  to semiclassical dynamics}},\ }\href {https://doi.org/10.1063/1.430620}
  {\bibfield  {journal} {\bibinfo  {journal} {The Journal of Chemical Physics}\
  }\textbf {\bibinfo {volume} {62}},\ \bibinfo {pages} {1544} (\bibinfo {year}
  {1975})}\BibitemShut {NoStop}%
\bibitem [{\citenamefont {Delande}(1991)}]{Delande}%
  \BibitemOpen
  \bibfield  {author} {\bibinfo {author} {\bibfnamefont {D.}~\bibnamefont
  {Delande}},\ }\href@noop {} {\emph {\bibinfo {title} {{Chaos and quantum
  physics, Les Houches Summer School, ed. MJ Giannoni, A Voros and J
  Zinn-Justin, Session LII}}}}\ (\bibinfo  {publisher} {North Holland,
  Amsterdam},\ \bibinfo {year} {1991})\BibitemShut {NoStop}%
\bibitem [{\citenamefont {Bluemel}\ and\ \citenamefont
  {Reinhardt}(1997)}]{Bluemel}%
  \BibitemOpen
  \bibfield  {author} {\bibinfo {author} {\bibfnamefont {R.}~\bibnamefont
  {Bluemel}}\ and\ \bibinfo {author} {\bibfnamefont {W.~P.}\ \bibnamefont
  {Reinhardt}},\ }\href@noop {} {\emph {\bibinfo {title} {Chaos in Atomic
  Physics}}}\ (\bibinfo  {publisher} {Cambridge University Press, Cambridge},\
  \bibinfo {year} {1997})\BibitemShut {NoStop}%
\bibitem [{\citenamefont {Zakrzewski}\ and\ \citenamefont
  {Delande}(1993)}]{Zakrzewski1993}%
  \BibitemOpen
  \bibfield  {author} {\bibinfo {author} {\bibfnamefont {J.}~\bibnamefont
  {Zakrzewski}}\ and\ \bibinfo {author} {\bibfnamefont {D.}~\bibnamefont
  {Delande}},\ }\bibfield  {title} {\bibinfo {title} {{Parametric motion of
  energy levels in quantum chaotic systems. I. Curvature distributions}},\
  }\href@noop {} {\bibfield  {journal} {\bibinfo  {journal} {Phys. Rev. E}\
  }\textbf {\bibinfo {volume} {47}},\ \bibinfo {pages} {1650} (\bibinfo {year}
  {1993})}\BibitemShut {NoStop}%
\bibitem [{\citenamefont {Zakrzewski}\ \emph {et~al.}(1993)\citenamefont
  {Zakrzewski}, \citenamefont {Delande},\ and\ \citenamefont
  {Ku{\'s}}}]{Zakrzewski1993b}%
  \BibitemOpen
  \bibfield  {author} {\bibinfo {author} {\bibfnamefont {J.}~\bibnamefont
  {Zakrzewski}}, \bibinfo {author} {\bibfnamefont {D.}~\bibnamefont
  {Delande}},\ and\ \bibinfo {author} {\bibfnamefont {M.}~\bibnamefont
  {Ku{\'s}}},\ }\bibfield  {title} {\bibinfo {title} {{Parametric motion of
  energy levels in quantum chaotic systems. II. Avoided-crossing
  distributions}},\ }\href@noop {} {\bibfield  {journal} {\bibinfo  {journal}
  {Phys. Rev. E}\ }\textbf {\bibinfo {volume} {47}},\ \bibinfo {pages} {1665}
  (\bibinfo {year} {1993})}\BibitemShut {NoStop}%
\bibitem [{\citenamefont {Weidenm{\"u}ller}\ and\ \citenamefont
  {Mitchell}(2009)}]{Weidenmuller2009}%
  \BibitemOpen
  \bibfield  {author} {\bibinfo {author} {\bibfnamefont {H.}~\bibnamefont
  {Weidenm{\"u}ller}}\ and\ \bibinfo {author} {\bibfnamefont {G.}~\bibnamefont
  {Mitchell}},\ }\bibfield  {title} {\bibinfo {title} {{Random matrices and
  chaos in nuclear physics: Nuclear structure}},\ }\href@noop {} {\bibfield
  {journal} {\bibinfo  {journal} {Rev. Mod. Phys.}\ }\textbf {\bibinfo {volume}
  {81}},\ \bibinfo {pages} {539} (\bibinfo {year} {2009})}\BibitemShut
  {NoStop}%
\bibitem [{\citenamefont {Haake}(2010)}]{Haakebook}%
  \BibitemOpen
  \bibfield  {author} {\bibinfo {author} {\bibfnamefont {F.}~\bibnamefont
  {Haake}},\ }\href@noop {} {\emph {\bibinfo {title} {{Quantum Signatures of
  Chaos}}}}\ (\bibinfo  {publisher} {Springer, Berlin},\ \bibinfo {year}
  {2010})\BibitemShut {NoStop}%
\bibitem [{\citenamefont {Stoeckmann}(2007)}]{Stoeckmann}%
  \BibitemOpen
  \bibfield  {author} {\bibinfo {author} {\bibfnamefont {H.-J.}\ \bibnamefont
  {Stoeckmann}},\ }\href@noop {} {\emph {\bibinfo {title} {{Quantum Chaos: An
  Introduction}}}}\ (\bibinfo  {publisher} {Cambridge University Press,
  Cambridge},\ \bibinfo {year} {2007})\BibitemShut {NoStop}%
\bibitem [{\citenamefont {Rosenzweig}\ and\ \citenamefont
  {Porter}(1960)}]{Rosenzweig1960}%
  \BibitemOpen
  \bibfield  {author} {\bibinfo {author} {\bibfnamefont {N.}~\bibnamefont
  {Rosenzweig}}\ and\ \bibinfo {author} {\bibfnamefont {C.~E.}\ \bibnamefont
  {Porter}},\ }\bibfield  {title} {\bibinfo {title} {{" Repulsion of Energy
  Levels" in Complex Atomic Spectra}},\ }\href@noop {} {\bibfield  {journal}
  {\bibinfo  {journal} {Physical Review}\ }\textbf {\bibinfo {volume} {120}},\
  \bibinfo {pages} {1698} (\bibinfo {year} {1960})}\BibitemShut {NoStop}%
\bibitem [{\citenamefont {Berry}\ and\ \citenamefont
  {Robnik}(1984)}]{Berry1984}%
  \BibitemOpen
  \bibfield  {author} {\bibinfo {author} {\bibfnamefont {M.~V.}\ \bibnamefont
  {Berry}}\ and\ \bibinfo {author} {\bibfnamefont {M.}~\bibnamefont {Robnik}},\
  }\bibfield  {title} {\bibinfo {title} {{Semiclassical level spacings when
  regular and chaotic orbits coexist}},\ }\href@noop {} {\bibfield  {journal}
  {\bibinfo  {journal} {Journal of Physics A: Mathematical and General}\
  }\textbf {\bibinfo {volume} {17}},\ \bibinfo {pages} {2413} (\bibinfo {year}
  {1984})}\BibitemShut {NoStop}%
\bibitem [{\citenamefont {Prosen}\ and\ \citenamefont
  {Robnik}(1999)}]{Prosen1999}%
  \BibitemOpen
  \bibfield  {author} {\bibinfo {author} {\bibfnamefont {T.}~\bibnamefont
  {Prosen}}\ and\ \bibinfo {author} {\bibfnamefont {M.}~\bibnamefont
  {Robnik}},\ }\bibfield  {title} {\bibinfo {title} {{Intermediate statistics
  in the regime of mixed classical dynamics}},\ }\href@noop {} {\bibfield
  {journal} {\bibinfo  {journal} {Journal of Physics A: Mathematical and
  General}\ }\textbf {\bibinfo {volume} {32}},\ \bibinfo {pages} {1863}
  (\bibinfo {year} {1999})}\BibitemShut {NoStop}%
\bibitem [{\citenamefont {Batisti{\'c}}\ \emph {et~al.}(2013)\citenamefont
  {Batisti{\'c}}, \citenamefont {Manos},\ and\ \citenamefont
  {Robnik}}]{Batistic2013}%
  \BibitemOpen
  \bibfield  {author} {\bibinfo {author} {\bibfnamefont {B.}~\bibnamefont
  {Batisti{\'c}}}, \bibinfo {author} {\bibfnamefont {T.}~\bibnamefont
  {Manos}},\ and\ \bibinfo {author} {\bibfnamefont {M.}~\bibnamefont
  {Robnik}},\ }\bibfield  {title} {\bibinfo {title} {{The intermediate level
  statistics in dynamically localized chaotic eigenstates}},\ }\href@noop {}
  {\bibfield  {journal} {\bibinfo  {journal} {Europhys. Lett.}\ }\textbf
  {\bibinfo {volume} {102}},\ \bibinfo {pages} {50008} (\bibinfo {year}
  {2013})}\BibitemShut {NoStop}%
\bibitem [{\citenamefont {Batisti{\'c}}\ \emph {et~al.}(2019)\citenamefont
  {Batisti{\'c}}, \citenamefont {Lozej},\ and\ \citenamefont
  {Robnik}}]{Batistic2019}%
  \BibitemOpen
  \bibfield  {author} {\bibinfo {author} {\bibfnamefont {B.}~\bibnamefont
  {Batisti{\'c}}}, \bibinfo {author} {\bibfnamefont {{\v{C}}.}~\bibnamefont
  {Lozej}},\ and\ \bibinfo {author} {\bibfnamefont {M.}~\bibnamefont
  {Robnik}},\ }\bibfield  {title} {\bibinfo {title} {{Statistical properties of
  the localization measure of chaotic eigenstates and the spectral statistics
  in a mixed-type billiard}},\ }\href@noop {} {\bibfield  {journal} {\bibinfo
  {journal} {Physical Review E}\ }\textbf {\bibinfo {volume} {100}},\ \bibinfo
  {pages} {062208} (\bibinfo {year} {2019})}\BibitemShut {NoStop}%
\bibitem [{\citenamefont {Kosicki}\ \emph {et~al.}(2020)\citenamefont
  {Kosicki}, \citenamefont {Borkowski},\ and\ \citenamefont
  {{\.{Z}}uchowski}}]{Kosicki2020}%
  \BibitemOpen
  \bibfield  {author} {\bibinfo {author} {\bibfnamefont {M.~B.}\ \bibnamefont
  {Kosicki}}, \bibinfo {author} {\bibfnamefont {M.}~\bibnamefont {Borkowski}},\
  and\ \bibinfo {author} {\bibfnamefont {P.~S.}\ \bibnamefont
  {{\.{Z}}uchowski}},\ }\bibfield  {title} {\bibinfo {title} {{Quantum chaos in
  Feshbach resonances of the {ErYb} system}},\ }\href
  {https://dx.doi.org/10.1088/1367-2630/ab6c36} {\bibfield  {journal} {\bibinfo
   {journal} {New J. Phys.}\ }\textbf {\bibinfo {volume} {22}},\ \bibinfo
  {pages} {023024} (\bibinfo {year} {2020})}\BibitemShut {NoStop}%
\bibitem [{\citenamefont {Gubin}\ and\ \citenamefont
  {F~Santos}(2012)}]{Gubin2012}%
  \BibitemOpen
  \bibfield  {author} {\bibinfo {author} {\bibfnamefont {A.}~\bibnamefont
  {Gubin}}\ and\ \bibinfo {author} {\bibfnamefont {L.}~\bibnamefont
  {F~Santos}},\ }\bibfield  {title} {\bibinfo {title} {{Quantum chaos: An
  introduction via chains of interacting spins 1/2}},\ }\href@noop {}
  {\bibfield  {journal} {\bibinfo  {journal} {American Journal of Physics}\
  }\textbf {\bibinfo {volume} {80}},\ \bibinfo {pages} {246} (\bibinfo {year}
  {2012})}\BibitemShut {NoStop}%
\bibitem [{\citenamefont {Giraud}\ \emph {et~al.}(2022)\citenamefont {Giraud},
  \citenamefont {Mac{\'e}}, \citenamefont {Vernier},\ and\ \citenamefont
  {Alet}}]{Giraud2022}%
  \BibitemOpen
  \bibfield  {author} {\bibinfo {author} {\bibfnamefont {O.}~\bibnamefont
  {Giraud}}, \bibinfo {author} {\bibfnamefont {N.}~\bibnamefont {Mac{\'e}}},
  \bibinfo {author} {\bibfnamefont {{\'E}.}~\bibnamefont {Vernier}},\ and\
  \bibinfo {author} {\bibfnamefont {F.}~\bibnamefont {Alet}},\ }\bibfield
  {title} {\bibinfo {title} {{Probing symmetries of quantum many-body systems
  through gap ratio statistics}},\ }\href@noop {} {\bibfield  {journal}
  {\bibinfo  {journal} {Phys. Rev. X}\ }\textbf {\bibinfo {volume} {12}},\
  \bibinfo {pages} {011006} (\bibinfo {year} {2022})}\BibitemShut {NoStop}%
\bibitem [{\citenamefont {Zakrzewski}(2023)}]{Zakrzewski2023}%
  \BibitemOpen
  \bibfield  {author} {\bibinfo {author} {\bibfnamefont {J.}~\bibnamefont
  {Zakrzewski}},\ }\bibfield  {title} {\bibinfo {title} {{Quantum chaos and
  level dynamics}},\ }\href@noop {} {\bibfield  {journal} {\bibinfo  {journal}
  {Entropy}\ }\textbf {\bibinfo {volume} {25}},\ \bibinfo {pages} {491}
  (\bibinfo {year} {2023})}\BibitemShut {NoStop}%
\bibitem [{\citenamefont {Yan}(2025)}]{Yan2025}%
  \BibitemOpen
  \bibfield  {author} {\bibinfo {author} {\bibfnamefont {H.}~\bibnamefont
  {Yan}},\ }\bibfield  {title} {\bibinfo {title} {{Spacing ratios in mixed-type
  systems}},\ }\href@noop {} {\bibfield  {journal} {\bibinfo  {journal} {Phys.
  Rev. E}\ }\textbf {\bibinfo {volume} {111}},\ \bibinfo {pages} {054213}
  (\bibinfo {year} {2025})}\BibitemShut {NoStop}%
\bibitem [{Sup()}]{SuppMat}%
  \BibitemOpen
  \href@noop {} {\bibinfo {title} {{See Supplemental Material for details on
  the Feshbach-spectrum measurement protocol, bound-state spectroscopy,
  additional robustness checks of the magnetic-moment-resolved level-statistics
  analysis, and a discussion of the complementarity between the Brody parameter
  $\beta$ and the average gap ratio $\langle r\rangle$.}}}\BibitemShut {Stop}%
\bibitem [{\citenamefont {Lucioni}\ \emph {et~al.}(2018)\citenamefont
  {Lucioni}, \citenamefont {Tanzi}, \citenamefont {Fregosi}, \citenamefont
  {Catani}, \citenamefont {Gozzini}, \citenamefont {Inguscio}, \citenamefont
  {Fioretti}, \citenamefont {Gabbanini},\ and\ \citenamefont
  {Modugno}}]{Lucioni2018}%
  \BibitemOpen
  \bibfield  {author} {\bibinfo {author} {\bibfnamefont {E.}~\bibnamefont
  {Lucioni}}, \bibinfo {author} {\bibfnamefont {L.}~\bibnamefont {Tanzi}},
  \bibinfo {author} {\bibfnamefont {A.}~\bibnamefont {Fregosi}}, \bibinfo
  {author} {\bibfnamefont {J.}~\bibnamefont {Catani}}, \bibinfo {author}
  {\bibfnamefont {S.}~\bibnamefont {Gozzini}}, \bibinfo {author} {\bibfnamefont
  {M.}~\bibnamefont {Inguscio}}, \bibinfo {author} {\bibfnamefont
  {A.}~\bibnamefont {Fioretti}}, \bibinfo {author} {\bibfnamefont
  {C.}~\bibnamefont {Gabbanini}},\ and\ \bibinfo {author} {\bibfnamefont
  {G.}~\bibnamefont {Modugno}},\ }\bibfield  {title} {\bibinfo {title}
  {{Dysprosium dipolar Bose-Einstein condensate with broad Feshbach
  resonances}},\ }\href {https://doi.org/10.1103/PhysRevA.97.060701} {\bibfield
   {journal} {\bibinfo  {journal} {Phys. Rev. A}\ }\textbf {\bibinfo {volume}
  {97}},\ \bibinfo {pages} {060701} (\bibinfo {year} {2018})}\BibitemShut
  {NoStop}%
\bibitem [{\citenamefont {Kao}\ \emph {et~al.}(2021)\citenamefont {Kao},
  \citenamefont {Li}, \citenamefont {Lin}, \citenamefont {Gopalakrishnan},\
  and\ \citenamefont {Lev}}]{Kao2021}%
  \BibitemOpen
  \bibfield  {author} {\bibinfo {author} {\bibfnamefont {W.}~\bibnamefont
  {Kao}}, \bibinfo {author} {\bibfnamefont {K.-Y.}\ \bibnamefont {Li}},
  \bibinfo {author} {\bibfnamefont {K.-Y.}\ \bibnamefont {Lin}}, \bibinfo
  {author} {\bibfnamefont {S.}~\bibnamefont {Gopalakrishnan}},\ and\ \bibinfo
  {author} {\bibfnamefont {B.~L.}\ \bibnamefont {Lev}},\ }\bibfield  {title}
  {\bibinfo {title} {{Topological pumping of a 1D dipolar gas into strongly
  correlated prethermal states}},\ }\href@noop {} {\bibfield  {journal}
  {\bibinfo  {journal} {Science}\ }\textbf {\bibinfo {volume} {371}},\ \bibinfo
  {pages} {296} (\bibinfo {year} {2021})}\BibitemShut {NoStop}%
\bibitem [{\citenamefont {Lecomte}\ \emph {et~al.}(2024)\citenamefont
  {Lecomte}, \citenamefont {Journeaux}, \citenamefont {Renaud}, \citenamefont
  {Dalibard},\ and\ \citenamefont {Lopes}}]{Lecomte2024}%
  \BibitemOpen
  \bibfield  {author} {\bibinfo {author} {\bibfnamefont {M.}~\bibnamefont
  {Lecomte}}, \bibinfo {author} {\bibfnamefont {A.}~\bibnamefont {Journeaux}},
  \bibinfo {author} {\bibfnamefont {L.}~\bibnamefont {Renaud}}, \bibinfo
  {author} {\bibfnamefont {J.}~\bibnamefont {Dalibard}},\ and\ \bibinfo
  {author} {\bibfnamefont {R.}~\bibnamefont {Lopes}},\ }\bibfield  {title}
  {\bibinfo {title} {{Loss features in ultracold $^{162}\mathrm{Dy}$ gases:
  Two- versus three-body processes}},\ }\href
  {https://doi.org/10.1103/PhysRevA.109.023319} {\bibfield  {journal} {\bibinfo
   {journal} {Phys. Rev. A}\ }\textbf {\bibinfo {volume} {109}},\ \bibinfo
  {pages} {023319} (\bibinfo {year} {2024})}\BibitemShut {NoStop}%
\bibitem [{\citenamefont {Duerbeck}\ \emph {et~al.}(2026)\citenamefont
  {Duerbeck}, \citenamefont {Reihs}, \citenamefont {Marulanda-Serna},
  \citenamefont {Choudhari}, \citenamefont {Seifert}, \citenamefont {Werum},
  \citenamefont {Meijer},\ and\ \citenamefont {Valtolina}}]{Duerbeck2026}%
  \BibitemOpen
  \bibfield  {author} {\bibinfo {author} {\bibfnamefont {M.}~\bibnamefont
  {Duerbeck}}, \bibinfo {author} {\bibfnamefont {L.}~\bibnamefont {Reihs}},
  \bibinfo {author} {\bibfnamefont {J.~P.}\ \bibnamefont {Marulanda-Serna}},
  \bibinfo {author} {\bibfnamefont {B.}~\bibnamefont {Choudhari}}, \bibinfo
  {author} {\bibfnamefont {J.}~\bibnamefont {Seifert}}, \bibinfo {author}
  {\bibfnamefont {N.}~\bibnamefont {Werum}}, \bibinfo {author} {\bibfnamefont
  {G.}~\bibnamefont {Meijer}},\ and\ \bibinfo {author} {\bibfnamefont
  {G.}~\bibnamefont {Valtolina}},\ }\href {https://arxiv.org/abs/2606.23528}
  {\bibinfo {title} {{A dipolar Bose-Bose mixture of Dysprosium isotopes with
  controllable interspecies interactions}}} (\bibinfo {year} {2026}),\ \Eprint
  {https://arxiv.org/abs/2606.23528} {arXiv:2606.23528 [cond-mat.quant-gas]}
  \BibitemShut {NoStop}%
\bibitem [{\citenamefont {Journeaux}\ \emph
  {et~al.}(2026{\natexlab{a}})\citenamefont {Journeaux}, \citenamefont
  {Veschambre}, \citenamefont {Lecomte}, \citenamefont {Uzan}, \citenamefont
  {Dalibard}, \citenamefont {Werner}, \citenamefont {Petrov},\ and\
  \citenamefont {Lopes}}]{Journeaux2026}%
  \BibitemOpen
  \bibfield  {author} {\bibinfo {author} {\bibfnamefont {A.}~\bibnamefont
  {Journeaux}}, \bibinfo {author} {\bibfnamefont {J.}~\bibnamefont
  {Veschambre}}, \bibinfo {author} {\bibfnamefont {M.}~\bibnamefont {Lecomte}},
  \bibinfo {author} {\bibfnamefont {E.}~\bibnamefont {Uzan}}, \bibinfo {author}
  {\bibfnamefont {J.}~\bibnamefont {Dalibard}}, \bibinfo {author}
  {\bibfnamefont {F.}~\bibnamefont {Werner}}, \bibinfo {author} {\bibfnamefont
  {D.~S.}\ \bibnamefont {Petrov}},\ and\ \bibinfo {author} {\bibfnamefont
  {R.}~\bibnamefont {Lopes}},\ }\bibfield  {title} {\bibinfo {title} {{Two-body
  contact dynamics in a Bose gas near a Fano-Feshbach resonance}},\ }\href@noop
  {} {\bibfield  {journal} {\bibinfo  {journal} {Phys. Rev. Lett.}\ }\textbf
  {\bibinfo {volume} {136}},\ \bibinfo {pages} {083404} (\bibinfo {year}
  {2026}{\natexlab{a}})}\BibitemShut {NoStop}%
\bibitem [{Note1()}]{Note1}%
  \BibitemOpen
  \bibinfo {note} {In practice this corresponds to having \begin
  {equation}\alpha = \left [ \Gamma \protect \!\left (\protect \frac {\beta
  +2}{\beta +1}\right ) \right ]^{\beta +1}\ , \end {equation}where $\Gamma
  (x)$ is the gamma function.}\BibitemShut {Stop}%
\bibitem [{Note2()}]{Note2}%
  \BibitemOpen
  \bibinfo {note} {For completeness, we also test for a possible smooth
  magnetic-field dependence of the Brody parameter by writing $\beta _i=\beta
  _0+\beta _1\protect \bar B_i$, where $\protect \bar
  B_i=(B_{i+1}+B_i)/2$~\cite {Makrides2018,Maier2015}. The resulting fit gives
  $\beta _0=0.54(10)$ and $\beta _1=-0.06(37)~{\protect \rm G}^{-1}$,
  consistent with no measurable field dependence over the investigated
  range.}\BibitemShut {Stop}%
\bibitem [{\citenamefont {Casella}\ and\ \citenamefont
  {Berger}(2002)}]{CasellaBerger2002}%
  \BibitemOpen
  \bibfield  {author} {\bibinfo {author} {\bibfnamefont {G.}~\bibnamefont
  {Casella}}\ and\ \bibinfo {author} {\bibfnamefont {R.~L.}\ \bibnamefont
  {Berger}},\ }\href@noop {} {\emph {\bibinfo {title} {{Statistical
  Inference}}}},\ \bibinfo {edition} {2nd}\ ed.\ (\bibinfo  {publisher}
  {Duxbury},\ \bibinfo {address} {Pacific Grove},\ \bibinfo {year}
  {2002})\BibitemShut {NoStop}%
\bibitem [{Note3()}]{Note3}%
  \BibitemOpen
  \bibinfo {note} {$\delta \mu _c$ is defined as the mean value of $\delta \mu
  $ over the selected resonances.}\BibitemShut {Stop}%
\bibitem [{\citenamefont {Oganesyan}\ and\ \citenamefont
  {Huse}(2007)}]{Oganesyan2007}%
  \BibitemOpen
  \bibfield  {author} {\bibinfo {author} {\bibfnamefont {V.}~\bibnamefont
  {Oganesyan}}\ and\ \bibinfo {author} {\bibfnamefont {D.~A.}\ \bibnamefont
  {Huse}},\ }\bibfield  {title} {\bibinfo {title} {{Localization of interacting
  fermions at high temperature}},\ }\href@noop {} {\bibfield  {journal}
  {\bibinfo  {journal} {Phys. Rev. B}\ }\textbf {\bibinfo {volume} {75}},\
  \bibinfo {pages} {155111} (\bibinfo {year} {2007})}\BibitemShut {NoStop}%
\bibitem [{\citenamefont {Atas}\ \emph {et~al.}(2013)\citenamefont {Atas},
  \citenamefont {Bogomolny}, \citenamefont {Giraud},\ and\ \citenamefont
  {Roux}}]{Atas2013}%
  \BibitemOpen
  \bibfield  {author} {\bibinfo {author} {\bibfnamefont {Y.~Y.}\ \bibnamefont
  {Atas}}, \bibinfo {author} {\bibfnamefont {E.}~\bibnamefont {Bogomolny}},
  \bibinfo {author} {\bibfnamefont {O.}~\bibnamefont {Giraud}},\ and\ \bibinfo
  {author} {\bibfnamefont {G.}~\bibnamefont {Roux}},\ }\bibfield  {title}
  {\bibinfo {title} {{Distribution of the ratio of consecutive level spacings
  in random matrix ensembles}},\ }\href@noop {} {\bibfield  {journal} {\bibinfo
   {journal} {Phys. Rev. Lett.}\ }\textbf {\bibinfo {volume} {110}},\ \bibinfo
  {pages} {084101} (\bibinfo {year} {2013})}\BibitemShut {NoStop}%
\bibitem [{\citenamefont {Journeaux}\ \emph
  {et~al.}(2026{\natexlab{b}})\citenamefont {Journeaux}, \citenamefont
  {Lecomte}, \citenamefont {Veschambre}, \citenamefont {Lepers}, \citenamefont
  {Dalibard},\ and\ \citenamefont {Lopes}}]{Journeaux2026b}%
  \BibitemOpen
  \bibfield  {author} {\bibinfo {author} {\bibfnamefont {A.}~\bibnamefont
  {Journeaux}}, \bibinfo {author} {\bibfnamefont {M.}~\bibnamefont {Lecomte}},
  \bibinfo {author} {\bibfnamefont {J.}~\bibnamefont {Veschambre}}, \bibinfo
  {author} {\bibfnamefont {M.}~\bibnamefont {Lepers}}, \bibinfo {author}
  {\bibfnamefont {J.}~\bibnamefont {Dalibard}},\ and\ \bibinfo {author}
  {\bibfnamefont {R.}~\bibnamefont {Lopes}},\ }\bibfield  {title} {\bibinfo
  {title} {{Determination of the ground state polarizability of $^{162}$ Dy
  near 530 nm}},\ }\href@noop {} {\bibfield  {journal} {\bibinfo  {journal}
  {arXiv:2604.03177}\ } (\bibinfo {year} {2026}{\natexlab{b}})}\BibitemShut
  {NoStop}%
\bibitem [{Note4()}]{Note4}%
  \BibitemOpen
  \bibinfo {note} {The same modulation technique is also used to calibrate the
  magnetic field and its dynamical response by driving the atomic transition
  $\protect \ensuremath {\left \vert -8\right \rangle }\to \protect \ensuremath
  {\left \vert -7\right \rangle }$. In this case, the observed atom loss
  results from dipolar relaxation involving atoms transferred to $\protect
  \ensuremath {\left \vert -7\right \rangle }$.}\BibitemShut {Stop}%
\bibitem [{\citenamefont {Chin}\ \emph {et~al.}(2010)\citenamefont {Chin},
  \citenamefont {Grimm}, \citenamefont {Julienne},\ and\ \citenamefont
  {Tiesinga}}]{ChinRMP10}%
  \BibitemOpen
  \bibfield  {author} {\bibinfo {author} {\bibfnamefont {C.}~\bibnamefont
  {Chin}}, \bibinfo {author} {\bibfnamefont {R.}~\bibnamefont {Grimm}},
  \bibinfo {author} {\bibfnamefont {P.~S.}\ \bibnamefont {Julienne}},\ and\
  \bibinfo {author} {\bibfnamefont {E.}~\bibnamefont {Tiesinga}},\ }\bibfield
  {title} {\bibinfo {title} {{Feshbach resonances in ultracold gases}},\ }\href
  {https://doi.org/10.1103/RevModPhys.82.1225} {\bibfield  {journal} {\bibinfo
  {journal} {Rev. Mod. Phys.}\ }\textbf {\bibinfo {volume} {82}},\ \bibinfo
  {pages} {1225} (\bibinfo {year} {2010})}\BibitemShut {NoStop}%
\bibitem [{\citenamefont {Veschambre}\ \emph {et~al.}()\citenamefont
  {Veschambre} \emph {et~al.}}]{Veschambre2026}%
  \BibitemOpen
  \bibfield  {author} {\bibinfo {author} {\bibfnamefont {J.}~\bibnamefont
  {Veschambre}} \emph {et~al.},\ }\href@noop {} {\bibinfo {title} {Article in
  preparation}}\BibitemShut {NoStop}%
\end{thebibliography}%

\cleardoublepage
\setcounter{equation}{0}
\setcounter{figure}{0}
\renewcommand{\theequation}{S\arabic{equation}}
\renewcommand{\thefigure}{S\arabic{figure}}

\appendix
\onecolumngrid 

\begin{center}
{\large \textbf{Supplemental Material for:}}\\[0.5em]
{\large \textbf{Coexisting Regular and Chaotic Dynamics in the Dysprosium Feshbach Spectrum
}}
\end{center}

\twocolumngrid

\section{Feshbach-spectrum measurement protocol}

For the measurements of the Feshbach spectrum, we start from non-degenerate spin-polarized samples of typically $10^5$ atoms at a temperature of about $350~{\rm nK}$, confined in an optical dipole trap with trap frequencies
$\pc{f_x,f_y,f_z}=\pc{32,281,229}~{\rm Hz}$, corresponding to a peak phase-space density of 0.5.
The final stage of evaporation is performed over $0.55~{\rm s}$ at a magnetic field of $1.52~{\rm G}$.
After evaporation, the magnetic field is quenched to a bias field $B_{\rm bias}$ chosen for the magnetic-field window under investigation.
This bias field is chosen sufficiently far from any observed Feshbach resonance, so that the atoms can be held without significant resonant loss while the magnetic field is stabilized.
More specifically, we use $B_{\rm bias}=2.3~{\rm G}$ for the window $B\in\pr{0,5.2}~{\rm G}$, $B_{\rm bias}=5.9~{\rm G}$ for $B\in\pr{5.2,11.9}~{\rm G}$, $B_{\rm bias}=11.9~{\rm G}$ for $B\in\pr{11.9,17.9}~{\rm G}$, $B_{\rm bias}=17.9~{\rm G}$ for $B\in\pr{17.9,20.2}~{\rm G}$, and $B_{\rm bias}=24.3~{\rm G}$ for $B\in\pr{20.2,30.4}~{\rm G}$.

\begin{figure*}[t!]
    \centering
    \includegraphics[width=15cm]{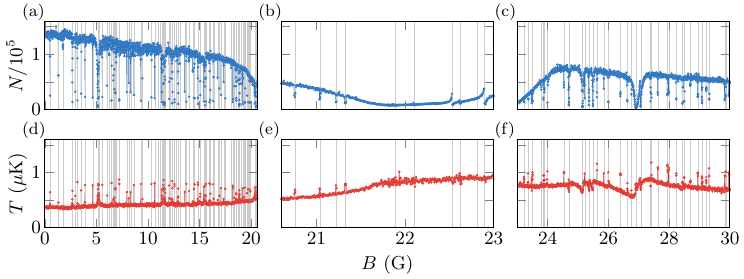}
    \caption{
    Atom number and temperature measured during the Feshbach-spectrum scan.
    Panels (a)--(c) show the remaining atom number as a function of magnetic field $B$.
    Panels (a) and (c) correspond to a hold time of $500~{\rm ms}$ at the final magnetic field, while panel (b) corresponds to a shorter hold time of $200~{\rm ms}$.
    Panels (d)--(f) show the corresponding temperature as a function of $B$, with the same hold times: $500~{\rm ms}$ for panels (d) and (f), and $200~{\rm ms}$ for panel (e).
    Vertical gray lines indicate the extracted positions of the Feshbach-resonance poles.
    In the vicinity of the resonance near $22~{\rm G}$, the atom-number signal displays asymmetric Fano-like profiles, visible in panel (b).
    }
    \label{figS1}
\end{figure*}

After reaching $B_{\rm bias}$, we hold the atoms for $500~{\rm ms}$ to allow the magnetic field to settle before starting the scan.
We then apply a linear magnetic-field ramp from $B_{\rm bias}$ to the target field $B$ at a rate $\dot B = 120~{\rm mG}/{\rm ms}$.
Once the target field is reached, the cloud is held for $500~{\rm ms}$ before the remaining atom number is measured by absorption imaging after time of flight. This hold time is used for all magnetic-field regions except around the broad loss feature near $22~{\rm G}$, where a shorter hold time of $200~{\rm ms}$ is used because of the strong atom losses associated with this resonance.
The Feshbach spectrum is obtained by repeating this sequence for target magnetic fields spaced by $6~{\rm mG}$ and recording the remaining atom number as a function of magnetic field. 

As the atoms are ramped to the target magnetic field, they may cross several Feshbach resonances before the hold time starts.
This can lead to additional heating and atom loss during the preparation sequence.
Consequently, the temperature measured after the hold time varies with the final magnetic field, as shown in Fig.~\ref{figS1}.
This effect is particularly visible in magnetic-field regions where the density of loss features is large.

Around $22~{\rm G}$, we also observe loss profiles with an asymmetric shape reminiscent of Fano resonances (see Fig.~\ref{figS1}b).
This behavior suggests that, in this region, the atomic sample is sensitive to two nearby molecular channels.
Interference between the corresponding loss pathways can then produce both enhanced and suppressed loss, leading to the asymmetric line shapes characteristic of Fano-like profiles.
Similar features have already been observed near the Fano resonance located at approximately $27~{\rm G}$ in the Supplemental Material of Ref.~\cite{Kao2021}, and, in the case where one pathway is associated with dipolar relaxation, in related work on Er~\cite{Lafforgue2025}.

\section{Bound-state spectroscopy}

We determine the energies of near-threshold molecular states using the spin-dependent light shift introduced in Refs.~\cite{Lecomte2025,Journeaux2026}. 
We use a laser beam with $1/e^2$ radius $w~=~95(2)~\mu\rm{m}$, and with frequency detuned by approximately $40~{\rm GHz}$ from the transition at $530.31~{\rm nm}$~\cite{Journeaux2026b}.
It is linearly polarized along the magnetic-field orientation and therefore couples the ground to the excited state manifold with  selection rule $\Delta m = 0$. Since the relevant excited manifold has angular momentum $J'=J-1=7$, atoms in the spin-polarized state $\ket{-8}$ are dark with respect to this transition, while other spin components experience a nonzero light shift. The entrance channel, composed of two atoms in $\ket{-8}$, is therefore mostly insensitive to this optical potential.

We modulate the intensity of this laser beam in time,
\begin{equation}
    I(t)=I_1\sin^2\left(\frac{\Omega t}{2}\right),
\end{equation}
which produces a time-dependent, spin-dependent energy shift at frequency $\Omega$. For the small peak intensity and large detuning used here, the resulting displacement of the Feshbach-resonance poles remains smaller than the magnetic-field step size of the scan.
This modulation is analogous to magnetic-field modulation, but can be implemented over a much broader frequency range. 
In practice, we observe spectroscopic resonances over the range $E/h=-21~{\rm MHz}$ to $E/h=6~{\rm MHz}$, for some bound-states.

When the modulation frequency matches the energy difference between the entrance-channel threshold and a molecular state, free atoms are coupled to the corresponding molecular state.
The resonance is detected as enhanced atom loss.
This loss can arise either from decay of the populated molecular state through three-body recombination or from photodissociation induced by the laser used for the modulation~\cite{Journeaux2026}.
By repeating this measurement at different magnetic fields, we reconstruct the near-threshold dispersion $E(B)$ of the molecular state.
\footnote{The same modulation technique is also used to calibrate the magnetic field and its dynamical response by driving the atomic transition $\ket{-8}\to\ket{-7}$.
In this case, the observed atom loss results from dipolar relaxation involving atoms transferred to $\ket{-7}$.}

For each molecular branch, we extract the differential magnetic moment from the local magnetic-field dependence of the molecular energy.
For the resonances considered here, the dispersion is well approximated over the relevant energy range by a linear relation,
\begin{equation}
    E(B)=\delta\mu\,(B-B_i),
\end{equation}
where $B_i$ is the magnetic field at which the molecular state is resonant with the entrance-channel threshold.
The slope $\delta\mu$ is the differential magnetic moment between the molecular state and the entrance-channel atom pair.

For all resonances included in the magnetic-moment analysis, the associated molecular feature is observed at least down to $E/h=-200~{\rm kHz}$ below the dissociation threshold.
This provides a common near-threshold reference for determining the local slope of each branch.
For some of the narrow resonances, the corresponding closed-channel molecular state can also be detected on the positive-energy side of the open-channel threshold.
Because the coupling to the entrance-channel continuum is weak, the state acquires only a small decay width and remains spectroscopically well defined even above threshold.
It can therefore be observed as a quasi-bound state embedded in the continuum~\cite{ChinRMP10}.
Whenever the signal remains visible at larger binding energies, additional spectroscopic points are used to improve the determination of $\delta\mu$.
For some resonances, as shown in Fig.~\ref{figFFRscan} of the main text, the same molecular branch can be followed down to -3~MHz and up to 3~MHz.

\section{Robustness of the magnetic-moment-resolved level-statistics analysis}

We test the robustness of the magnetic-moment-resolved level-statistics analysis using a sliding-window procedure in our selection of differential magnetic moments. 
Each resonance is characterized by its magnetic-field position $B_i$ and by its measured differential magnetic moment $\delta\mu_i$, with uncertainty $\sigma_{\delta\mu_i}$. The analysis presented in the main text uses the measured central values of $\delta\mu_i$. 
Here, as an additional robustness check, we propagate the experimental uncertainties on the measured $\delta\mu_i$ values into the extracted $\beta(\delta\mu_c)$ and $\langle r\rangle(\delta\mu_c)$. 
This is done by repeating the analysis over Monte-Carlo realizations in which each $\delta\mu_i$ is randomly drawn from a normal distribution centered on its measured value, with a standard deviation given by its experimental uncertainty. As shown here, this additional resampling does not qualitatively modify the results.

For each realization, resonances are grouped into sliding windows in $\delta\mu$. 
The window boundaries are adjusted so that each window contains the same number of Feshbach resonances, $N_{\rm res}$. 
Because the scan step is smaller than the typical window width, neighboring windows overlap and share resonances. This differs from the representative subsets shown in the main text, which are chosen to be non-overlapping. 
Here, the sliding-window procedure is used only to test the robustness and continuity of the magnetic-moment dependence of the level statistics.
The central magnetic moment $\delta\mu_c$ of each window is defined as the mean differential magnetic moment of the resonances it contains.

\begin{figure}[t!]
    \centering
    \includegraphics[width=8cm]{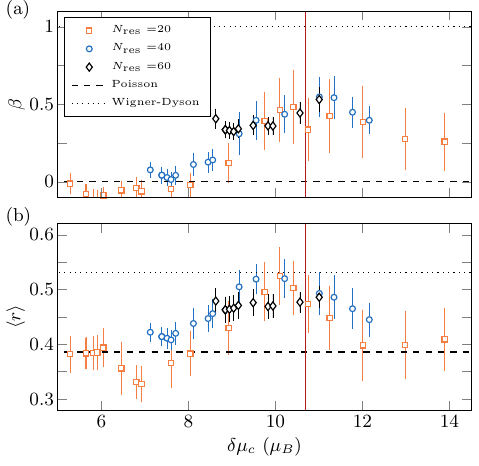}
\caption{
Dependence of the spectral statistics on the central differential magnetic moment $\delta\mu_c$ of the selected resonance subset. 
(a) Brody parameter $\beta$ as a function of $\delta\mu_c$ for selections containing $N_{\rm res}=20$, $40$, and $60$ resonances, shown as orange squares, blue circles, and black diamonds, respectively. 
(b) Average ratio of consecutive level spacings $\langle r\rangle$ as a function of $\delta\mu_c$ for the same values of $N_{\rm res}$. 
The selections in $\delta\mu$ are performed after taking into account the uncertainty of each individual measured $\delta\mu$ value, as described in the text. 
The vertical red line indicates the mean differential magnetic moment of the full distribution, $\overline{\delta\mu}$. 
In both panels, the dashed and dotted horizontal lines indicate the Poisson and GOE limits, respectively.
}
    \label{figS2}
\end{figure}

For each selected subset, the resonance positions are unfolded independently before extracting the Brody parameter. 
The accumulated number of resonances is fitted by a smooth second-order polynomial, and the unfolded spacings are obtained from the fitted unfolding function. 
The Brody parameter $\beta$ is then determined from the individual unfolded spacings using the same unbinned maximum-likelihood procedure as in the main text. 
This produces a value of $\beta$ as a function of the central differential magnetic moment, $\beta(\delta\mu_c)$.

In Fig.~\ref{figS2}(a), we show the evolution of $\beta$ as a function of $\delta\mu_c$ for three fixed-count window sizes, $N_{\rm res}=20$, $40$, and $60$. 
For windows centered near the mean of the differential-magnetic-moment distribution, $\delta\mu_c\simeq\overline{\delta\mu}=10.70 (13)\,\mu_{\rm B}$, we consistently find a relatively large Brody parameter, $\beta\simeq 0.6$. 
Near the lower edge of the $\delta\mu$ distribution, the result depends more strongly on the window size. 
For sufficiently selective windows, corresponding to smaller $N_{\rm res}$, the extracted $\beta$ is consistent with zero, indicating statistics close to the Poisson limit. 
As $N_{\rm res}$ is increased, the window necessarily includes a larger fraction of resonances from the central part of the $\delta\mu$ distribution, and the extracted value of $\beta$ correspondingly increases. 
For the largest window shown, $N_{\rm res}=60$, the magnetic-moment dependence is therefore strongly averaged out and no significant variation of $\beta$ with $\delta\mu_c$ is resolved.
Toward the upper edge of the distribution, $\beta$ also appears to decrease, although this region should be interpreted with more care. 
Both experimental and physical considerations may contribute to this behavior. 
Experimentally, resonances associated with states having a strong magnetic-field dependence become harder to detect with finite magnetic-field resolution. 
Physically, these states are expected to involve channels with larger orbital angular momentum, for which centrifugal barriers reduce the density of accessible near-threshold bound states (see main text).

In defining the representative low-moment subset shown in Fig.~\ref{figBrodyMu} of the main text, we exclude the resonance near $22~{\rm G}$ with $\delta\mu\simeq0.1\,\mu_{\rm B}$. 
This resonance is associated with a distinct, strongly open-channel-dominated feature and will be discussed separately~\cite{Veschambre2026}. 
The sliding-window analysis shown in Fig.~\ref{figS2} confirms that the magnetic-moment dependence of the level statistics does not rely on this particular choice.

We also perform the same sliding-window analysis for the mean ratio of consecutive spacings $\langle r\rangle$,
\begin{equation}
    r_i=
    \min\left(
    \frac{d_i}{d_{i-1}},
    \frac{d_{i-1}}{d_i}
    \right),
    \label{Eqri}
\end{equation}
where $d_i=B_{i+1}-B_i$ are the raw magnetic-field spacings within the selected subset. 
This quantity provides an unfolding-free check of the magnetic-moment dependence observed in the Brody analysis. 
The uncertainty on $\langle r\rangle$ is estimated by bootstrap resampling of the set of $r_i$ values. 
As shown in Fig.~\ref{figS2}(b), $\langle r\rangle$ follows a similar trend to $\beta$: it is larger near the center of the differential-magnetic-moment distribution and decreases toward values close to the Poisson expectation near the lower edge. 
This analysis confirms that, despite the finite size of the data set, the observed magnetic-moment dependence does not rely on a fine-tuned choice of selection window or statistical estimator, and remains robust within the experimental uncertainties.

\section{Complementarity of $\beta$ and $\av{r}$}

In the main text, we use both the Brody parameter $\beta$ and the average gap ratio $\av{r}$ to characterize the short-range spectral statistics. These two quantities are related, but they do not probe the same information. The Brody parameter is extracted from the nearest-neighbor spacing distribution $P(s)$, whereas $\av{r}$ depends on pairs of consecutive spacings, and thus probes the correlation between the two members of the pair. Therefore, $\beta$ and $\av{r}$ should be regarded as complementary diagnostics rather than interchangeable measures of the same quantity.

To illustrate this difference, we perform a simple exercise. We consider the case $\beta=1$, for which the Brody distribution reduces to the Wigner-surmise form
\be
P(s)=\frac{\pi}{2}s \exp\left(-\frac{\pi s^2}{4}\right) \ ,
\ee
we perform a random draw of $N$ values of $s_i$ for this statistical law, and we construct the distribution of $B_i$'s according to $B_0=0$, $B_{i+1}=B_i + s_{i+1}$ for $i=1, \ldots ,  N$. By construction, the set of $B_i$'s lead to the same $\beta$ as the GOE. On the other hand, for these independent drawn spacings, the average gap ratio then reduces to 
\be
\av{r} = 2\int_{0}^{1}\!{\rm d}r \ r\int_{0}^{\infty}\!{\rm d}t\ t\ P(rt)P(t) = \frac{\pi}{2}-1\simeq0.571.
\ee
This value is different from the GOE expectation, $\av{r}_{\rm GOE}\simeq0.531$~\cite{Atas2013, Giraud2022}.

\end{document}